\documentclass{cimento}
\usepackage{epsfig,amssymb}
\title{On the parametric approximation in quantum optics}
\author{G. M. D'Ariano, M. G. A. Paris and M. F. Sacchi}
\instlist{
\inst Dipartimento di Fisica \lq Alessandro Volta\rq and INFM---Unit\`a di
PAVIA,\\Universit\`a degli Studi di Pavia \\
via A. Bassi 6, I-27100 Pavia, Italy}
\PACSes{\PACSit{03.65.-w, 42.50.-p, 42.65.Yj, 42.50.Ar}{}}

\begin{document}

\maketitle

\begin{abstract}
We perform the exact numerical diagonalization of the Hamiltonians
that describe both degenerate and nondegenerate parametric amplifiers,
by exploiting the conservation laws pertaining each device. We clarify
the conditions under which the parametric approximation holds, showing
that the most relevant requirement is the coherence of the pump after
the interaction, rather than its undepletion.
\end{abstract}

\section{Introduction}
The interactions of different radiation modes through nonlinear
crystals allow the generation of interesting states of light, which
exhibit a rich variety of phenomena
\cite{mandel,per,rev,yue,yam,spe,iss,shih,ou,rarity,sac,dema,kly,horne,igo}.
Most of the theoretical analysis usually refers to situations where
one mode---the so-called ``pump'' mode---is placed in a high-amplitude
coherent state. In such cases the parametric approximation is widely
used to compute the dynamical evolution \cite{mol}.  In the parametric
approximation the pump mode is classically treated as a $c$-number,
thus neglecting both the depletion mechanism and quantum fluctuations.
As a result, bilinear and trilinear Hamiltonians are reduced to linear
and quadratic forms in the field operators, respectively, and hence
some useful mathematical tools---typically decomposition formulas for
Lie algebras---can be exploited for calculations \cite{rev,tru,su}.
\par In the validity regime of the parametric approximation different
optical devices experimentally realize different fundamental unitary
operators in quantum optics. For example, a beam splitter, by suitably
mixing the signal state with a strong local oscillator at the same
frequency, realizes the displacement operator, which generates
coherent states from the vacuum. Similarly, a degenerate parametric
amplifier realizes the squeezing operator, which is the generator of
the squeezed vacuum.  Finally, a nondegenerate parametric amplifier
realizes the two-mode squeezing operator, i.e. the generator of
twin-beam.  In the above scenario, it is matter of great interest to
study the conditions under which the parametric approximation holds.
This issue has been considered by a number of authors
\cite{bsparis,hizu,cro,scf}, however without giving a
general validity criterion, which is the main concern of this paper.
Quantum effects in two-mode optical amplifiers have been extensively
analyzed \cite{hizu,cro,bak,tin,buck,buz2,mos,tan,buz,hi2}.  Frequency
couplers with intensity dependent coupling have been studied
\cite{buck,buz2,dat}, whereas the case of degenerate parametric
amplifier has been considered by many authors
\cite{hizu,cro,tin,mos,tan,buz,hi2}.  Phase correlations \cite{tan}
and the signal-pump degree of entanglement \cite{buz} have been
examined.  The effect of pump squeezing has been also considered
\cite{hi2}.  On the other hand, though trilinear processes have been
thoroughly analyzed in Refs. \cite{igo,tin,igogab}, only little
attention has been devoted to the parametric approximation in
nondegenerate amplifiers \cite{scf}.  \par The most explicit
conditions for the validity of the parametric approximation have been
derived in Refs. \cite{bsparis} and \cite{hizu}, for the beam splitter
and the degenerate parametric amplifier, respectively.  In both
references sufficient conditions have been derived, which however can
be widely breached in relevant cases of interest, as we will show in
the following.  \par In this paper we perform the exact numerical
diagonalization of the full Hamiltonians pertaining the three
above-mentioned devices.  As it was already noted by other authors
\cite{bak,tin,igogab} such a numerical treatment is made amenable by
the presence of constants of motion that characterizes each
Hamiltonian . In fact, the Hilbert space can be decomposed into the
direct sum of subspaces that are invariant under the action of the
unitary evolution. Therefore, one needs to diagonalize the Hamiltonian
just inside each invariant subspace, thus considerably reducing the
dimension of the diagonalization space \cite{bak,tin,igogab}.  \par We
analyze in different sections the cases of the beam splitter, the
degenerate parametric amplifier and the nondegenerate parametric
amplifier.  The case of the beam splitter can be treated analytically,
but we also present some numerical results in order to introduce the
general approach that will be used for the parametric amplifiers.  For
each device, we look for the conditions under which the parametric
approximation holds, for both vacuum and non-vacuum input signal
states.  The comparison between the exact evolution and the
theoretical predictions from the parametric approximation is made in
terms of the overlap ${\cal O}=\sqrt
{\hbox{Tr}(\hat{\varrho}_{\hbox{\footnotesize{out}}}\;\hat\varrho
_{\hbox{\footnotesize{th}}} )}$ between the state $\hat\varrho
_{\hbox{\footnotesize{out}}}$ that exits the device and the
theoretical state $\hat\varrho_{\hbox {\footnotesize{th}}}$ obtained
within the approximation.  An explicit comparison in terms of photon
number distributions and Wigner functions is also given for some
interesting and representative cases.  \par The main result of the
paper is to show that the usual requirements for the validity of the
parametric approximation, namely short interaction time and strong
classical undepleted pump, are too strict.  Indeed, we show that the
only relevant requirement is the coherence of the pump after the
interaction, rather than its undepletion.  In fact, we will show
typical examples in which the pump at the input is weak (one photon in
average), after the interaction it is highly depleted, and
notwithstanding the parametric approximation still holds.  On the
other hand, there are cases in which the pump after the interaction is
only slightly depleted, however is no longer coherent, and the
approximation fails. Finally, we show some interesting features such
as pump squeezing and Schr\"odinger-cat-like state generation that
arise when the parametric approximation breaks down.
\section{Displacer}
The beam splitter is a passive device that couples two different modes 
of radiation at the same frequency through a first-order 
susceptibility-tensor $\chi ^{(1)}$ medium. 
Such device is widely used in quantum optics 
\cite{mandel,cam}, from homodyne/heterodyne detection 
\cite{yueopt}, to directional couplers \cite{jan} and cavity QED \cite{fea}. 
In the rotating wave approximation and under phase-matching conditions, 
the beam splitter Hamiltonian 
writes in terms of the two mode operators $a$ and $b$ as follows
\begin{equation}
\hat H _{BS} =\kappa \left(a b^{\dag}+a^{\dag }b\right)
\label{hbs}\;,
\end{equation}
where $\kappa $ is the coupling constant proportional to the 
$\chi ^{(1)}$ of the medium. 
The unitary evolution operator of the device in the interaction picture 
writes
\begin{eqnarray}
\hat U _{BS}=\exp\left[-i \tau\left(a b^{\dag}+a^{\dag }b\right)\right]
= e^{-i\tan\tau ab^{\dag}}\,
|\cos \tau|^{a^{\dag}a-b^\dag b}\,e^{-i\tan\tau a^{\dag} b}
\label{ubs}\;,
\end{eqnarray}
where $\tau $ is the interaction time rescaled by the coupling $\kappa $. 
The factorization of the operator $\hat U_{BS}$ in Eq. (\ref{ubs}) 
has been derived by applying the 
Baker-Campbell-Hausdorff formula for the SU(2) algebra \cite{rev,tru,su}. 
The Heisenberg evolution of the field modes reads
\begin{eqnarray}
\hat U_{BS}^{\dag} \,
\left( \begin{array}{c}a \\ b\end{array}\right)\,\hat U_{BS}
=\left( \begin{array}{c} 
a\,\cos\tau-ib\,\sin\tau\\ 
-ia\,\sin\tau+b\,\cos\tau
\end{array}\right) \;.\label{heisbs}
\end{eqnarray}
From Eq. (\ref{heisbs}) it turns out that 
the transmissivity $\theta$ at the beam splitter is given by the relation 
$\theta=\cos ^2 \tau$.  
\par The parametric approximation refers to situations in which one 
mode---say mode $b$---is excited in a strong coherent state. In this case 
in the first line of Eq. (\ref{ubs}) the operator $b$ might be replaced by 
a $c$-number, namely the complex amplitude $\beta$ of the coherent state. 
Under this assumption, the evolution operator (\ref{ubs}) would rewrite as 
the following displacement operator
\begin{equation}
\hat D(-i\beta\tau )\equiv\exp\left[-i\tau\left(\beta a^{\dag}+
\bar\beta a\right)\right]
\label{dispbad}\;.
\end{equation}
A more refined approximation that takes into account the 
$2\pi$-periodicity in the exact 
Heisenberg equations (\ref{heisbs}) is 
\begin{equation}
\hat D(-i\beta\sin\tau )\equiv\exp\left[-i \sin\tau\left(\beta a^{\dag}+
\bar\beta a\right)\right]
\label{disp}\;.
\end{equation}
Indeed, the more precise result in Eq. 
(\ref{disp}) can be obtained by recasting the factorized expression in 
Eq. (\ref{ubs}) in normal order with respect to mode $a$, after taking the 
expectation over mode $b$ \cite{nvcim,bilk}. 
The simple form of the bilinear Hamiltonian in Eq. (\ref{hbs}) allows 
to clarify the conditions under which the parametric 
approximation (\ref{disp}) holds \cite{bsparis}. A set of sufficient 
requirements are given by
\begin{eqnarray}
&&|\beta|\rightarrow \infty\;,\qquad \sin\tau\rightarrow 0\nonumber \\
&&|\beta|\sin\tau =\hbox{constant}\;, \label{condbs}
\end{eqnarray}
without any assumption on the ``signal'' state for mode $a$. 
Hence, by combining a signal input state $\hat\varrho 
_{\hbox{\footnotesize{in}}}$ with a strong coherent local 
oscillator $|\beta\rangle $ 
in a beam splitter with very high transmissivity, one can 
achieve the displacement operator in Eq. (\ref{disp}). The theoretically 
expected state $\hat\varrho 
_{\hbox{\footnotesize{th}}}$ then writes
\begin{equation}
\hat\varrho 
_{\hbox{\footnotesize{th}}}=\hat D(-i\beta\sin\tau)\,\hat\varrho 
_{\hbox{\footnotesize{in}}}\,\hat D^{\dag}(-i\beta\sin\tau)
\label{rth}\;.
\end{equation}
\par 
Here we present some numerical results concerning 
the exact unitary evolution of Eq. (\ref{ubs}). 
The dynamics generated by the Hamiltonian (\ref{hbs}) 
preserves the total number of photons involved in the process, 
in agreement with the following commutation relation
\begin{equation}
\left[\hat H _{BS},a^{\dag}a+b^{\dag}b\right]=0
\label{com}\;.
\end{equation}
Therefore, it is convenient to decompose the 
Hilbert space ${\cal H}_a \otimes {\cal H}_b$ as a direct sum of 
subspaces with a fixed number $N$ of photons, since these are invariant 
under the action of the unitary evolution operator (\ref{ubs}). 
Such a decomposition can be written as follows
\begin{eqnarray}
{\cal H}_a \otimes {\cal H}_b=\oplus _{N=0}^{+\infty} 
{\cal H}_N \label{decbs1}\;
\end{eqnarray}
where
\begin{eqnarray}
{\cal H}_N=\hbox{Span}\left\{
|m\rangle \otimes|N-m\rangle \;, m\in [0,N]\right\}
\label{decbs2}\;,
\end{eqnarray}
$\hbox{Span}\{\cdot\}$ denoting the Hilbert subspace linearly spanned by 
the orthogonal vectors within the brackets, and $\vert n\rangle \otimes 
\vert m\rangle \equiv \vert n,m\rangle$ representing the common 
eigenvector of the number operator of the two modes.
The decomposition in Eq. (\ref{decbs1}) makes the Hamiltonian (\ref{hbs}) 
block-diagonal, namely
\begin{eqnarray}
\hat H_{BS}=\sum _{N=0}^{+\infty}\hat h_N
\label{hn}\;,
\end{eqnarray}
where $\hat h_N$ acts just inside the subspace ${\cal H}_N$. 
Correspondingly, a generic two-mode state $\vert\psi_0\rangle$ can be 
written in the orthogonal basis (\ref{decbs2}) as follows
\begin{eqnarray}
\vert\psi_0\rangle= \sum_{N=0}^{+\infty} \sum_{m=0}^{N}
c_{m,N-m}\:\vert m,N-m\rangle
\label{spe}\;.
\end{eqnarray}
The diagonalization is performed inside each invariant subspace, and
the truncation of the series in Eqs. 
(\ref{hn}) and (\ref{spe}) corresponds to fix the maximum eigenvalue 
of the constant of motion $a^{\dag}a+b^{\dag}b$. 
\par The state $\hat\varrho _{\hbox{\footnotesize{out}}}$ 
evaluated by the exact evolution operator (\ref{ubs}) is given by
\begin{equation}
\hat\varrho _{\hbox{\footnotesize{out}}}=
\hbox{Tr}_{b}[\hat U_{BS}(\hat\varrho _{\hbox{\footnotesize{in}}}\otimes 
|\beta\rangle\langle\beta|)\hat U^{\dag}_{BS}]\;,
\end{equation}
where $\hbox{Tr}_{b}$ denotes the partial trace on ${\cal H}_b$. 
The comparison between the theoretical state 
$\hat\varrho_{\hbox{\footnotesize{th}}}$ of Eq. (\ref{rth}) within 
the parametric approximation and the actual state 
$\hat\varrho _{\hbox{\footnotesize{out}}}$ is 
made in terms of the relative overlap
\begin{equation}
{\cal O}\equiv\sqrt{\hbox{Tr}[\hat\varrho 
_{\hbox{\footnotesize{th}}}
\hat\varrho _{\hbox{\footnotesize{out}}}]}\;.\label{ovbs}
\end{equation}
In the case of coherent input signal the overlap is 
evaluated analytically. One has
\begin{eqnarray}
\hat U_{BS}\quad |\alpha\rangle\otimes|\beta\rangle= 
 |\alpha\cos\tau-i\beta\sin\tau\rangle\otimes|\beta\cos\tau-i\alpha
\sin\tau\rangle\;,
\end{eqnarray}
and thus
\begin{eqnarray}
{\cal O}=|\langle \alpha-i\beta\sin\tau|
\alpha\cos\tau-i\beta\sin\tau\rangle|= \exp\left(-4|\alpha |^2\sin
^4\frac{\tau}{2}\right)\label{ovth}\;.
\end{eqnarray}
From Eq. (\ref{ovth}) it is apparent that the parametric approximation 
gives always exact results for vacuum input state ($\alpha\equiv 0$), 
whereas it is justified for coherent state as long as 
$4|\alpha |^2\sin ^4(\tau /2)\ll 1$, independently on the pump intensity. 
\par
\begin{figure}
\begin{center}
\epsfxsize=.7\textwidth\leavevmode\epsffile{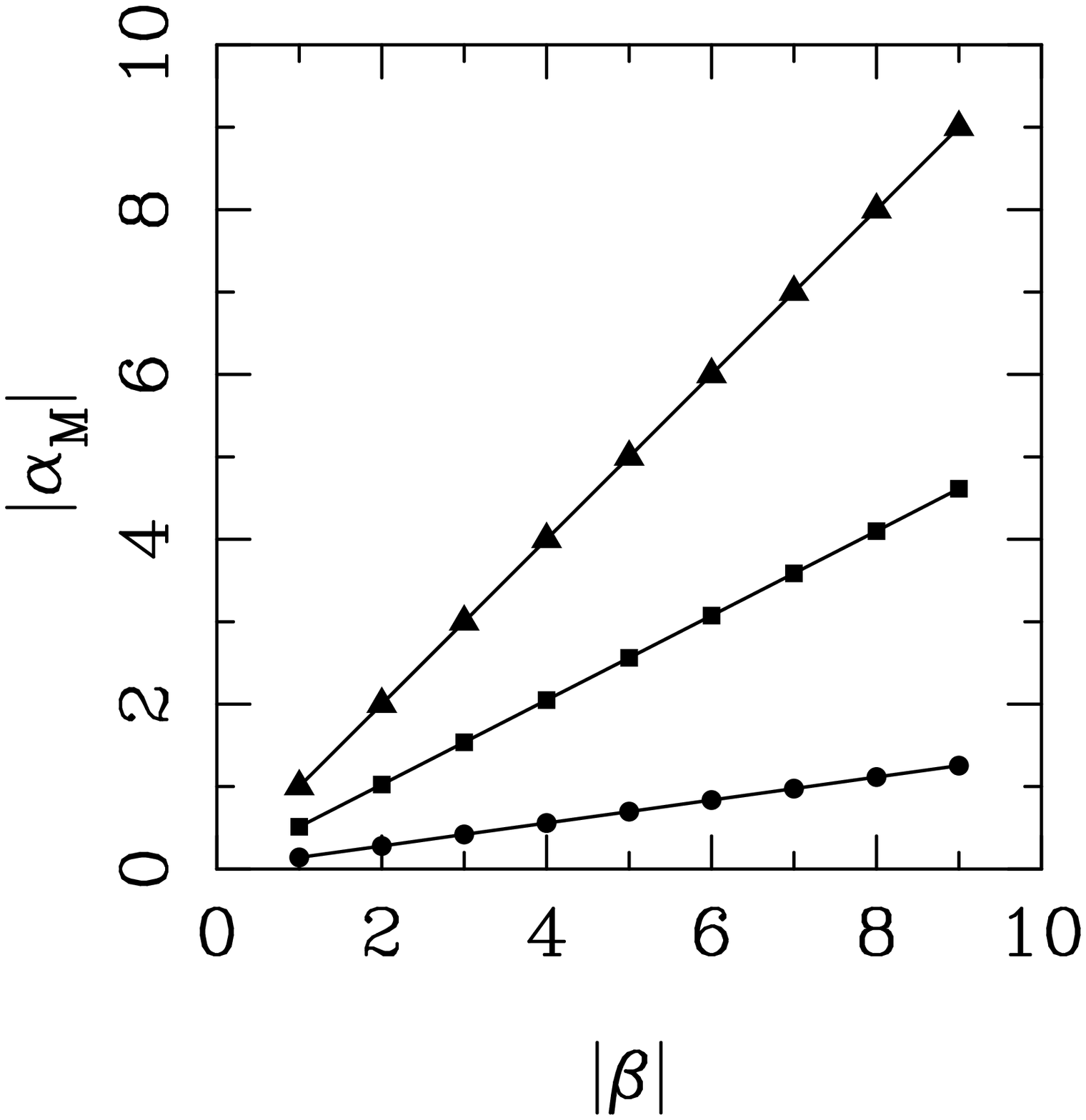}
\end{center}
\vskip .8cm
\caption{Performances of a beam splitter in achieving the displacement
operator. We report the maximum displacing amplitude $\vert z_{M}
\vert$ achievable by a beam splitter as a function of the pump
amplitude $|\beta|$. In the picture triangles refers to vacuum input,
squares to coherent state input $\vert\alpha\equiv 1\rangle$, and
circles to number state input $\vert n\equiv 1\rangle$.}
\label{f:bs1}
\end{figure}

We introduce the quantity $\tau^{\star}$ which represents, for a fixed 
value of the pump amplitude $|\beta|$, the maximum interaction time 
leading to an overlap larger than $99\%$. 
The value of $\tau^{\star}$ clearly depends on the input signal:
in agreement with Eq. (\ref{ovth}) it is not defined for the vacuum
(parametric approximation is exact), whereas for a coherent input 
signal $|\alpha\rangle$ it is given by
$$ 
\tau^\star = 2 \arcsin \sqrt{\frac{C}{\vert\alpha\vert}} \qquad 
C=\frac{1}{2}\left(-\ln 0.99 \right)^{1/2}\simeq 0.05
$$
The quantity $\tau^{\star}$ also determines the maximum displacing 
amplitude $|z_M|\equiv |\beta|\sin \tau^{\star}$ that can be 
achieved by a beam splitter with a coherent pump $|\beta\rangle$. 
In Fig. \ref{f:bs1} we have reported $|z_M|$ for the vacuum, a coherent state 
and a number state as a function of the pump amplitude $\vert\beta\vert$. 
The linear behavior of the plots indicates that  
$\tau^\star$ is independent on the pump intensity. In the case of vacuum input 
we have complete energy transfer from the pump to the signal 
(slope of $|z_M |$ vs $|\beta |$ equal to unit). 
Although they have the same energy, 
the coherent and number input states show different slopes, the 
coherent being more similar to the vacuum. 
Actually the set of coherent states is closed 
under the action of the displacement operator, so that 
the parametric approximation can fail only in predicting the exact 
amplitude of the output coherent state. \par 
We conclude that the first of requirements (\ref{condbs}) is too tight. 
At least for coherent and number states, as long as the signal average 
photon number is less than the pump one, the beam splitter can ``displace'' 
the signal also for very weak pump. 
\par
In the next sections we will deal with the problem of parametric 
approximation in nonlinear 
amplifiers.
\section{Squeezer}
The degenerate parametric amplifier couples a signal mode $a$ at 
frequency $\omega _a$ with a pump mode $c$ at double frequency 
$\omega _c=2\omega _a$. The interaction is mediated by the second-order 
susceptibility tensor  $\chi ^{(2)}$ of the medium. Each photon in the pump 
mode produces a photon pair in the signal mode, giving rise to light with a 
number of interesting properties, such as phase-sensitive amplification, 
squeezing and antibunching \cite{yue,yam,spe,iss,igo}. 
In the rotating wave approximation and under 
phase-matching conditions the Hamiltonian writes
\begin{eqnarray}
\hat H_{DP}=\kappa \left(a^2c^{\dag }+a^{\dag 2}c\right)\label{hdp}\;,
\end{eqnarray}
with $\kappa \propto \chi ^{(2)}$. The corresponding unitary evolution 
operator in the interaction picture reads 
\begin{eqnarray}
\hat U _{DP}=\exp\left[-i \tau\left(a^2 c^{\dag }+a^{\dag 2}c\right)\right]
\label{udp}\;,
\end{eqnarray}
where $\tau$ represents a rescaled interaction time. 
The parametric approximation replaces the pump mode $c$ by 
the complex amplitude $\beta$ of the corresponding coherent state, such that 
the operator (\ref{udp}) rewrites as 
\begin{eqnarray}
\hat S(-2i\tau\beta)\equiv \exp\left[-i\tau\left(\beta a^{\dag 2}+\bar\beta a^2
\right) \right]\;,\label{squez} 
\end{eqnarray}
$\hat S(\zeta )$ being the squeezing operator \cite{yue}.  
In the case of coherent input signal $|\alpha\rangle$, 
the predicted state at the output is the squeezed state
\begin{eqnarray}
\hat S(-2i\tau\beta)|\alpha\rangle =\hat S(-2i\tau\beta)\hat D(\alpha
) |0\rangle = \hat D(\tilde\alpha)\hat S(-2i\tau\beta)|0\rangle \equiv
|\tilde\alpha\,,-2i\tau\beta\rangle
\label{sqth}\;,
\end{eqnarray}
with $\tilde \alpha=\alpha\cosh (-2i\tau\beta)+\bar\alpha\sinh 
(-2i\tau\beta)$. 
Notice that, differently from the 
beam splitter operator of Eq. (\ref{ubs}), we have no method available 
to order Eq. (\ref{udp}) normally with respect to mode $c$ [as in Eq. 
(\ref{ubs}) for $b$] and then replace such mode by the $c$-number $|\beta |$. 
Hence, we have no analogous nonperturbative method to estimate the 
validity of the parametric approximation. 
Hillery and Zubairy have been approached the question \cite{hizu} 
in terms of a perturbation series for the propagator of the Hamiltonian 
(\ref{hdp}). For initial vacuum state at mode $a$, they write the 
following conditions
\begin{eqnarray}
&&1/|\beta|\ll 1\;, \qquad \tau \ll 1\;,\nonumber \\
&&\tau e^{4|\beta |\tau}\ll 1\;,\qquad e^{4|\beta|\tau}\ll |\beta|
\;.\label{conddp}
\end{eqnarray}
\par
\begin{figure}
\begin{center}
\epsfxsize=.49\textwidth\leavevmode\epsffile{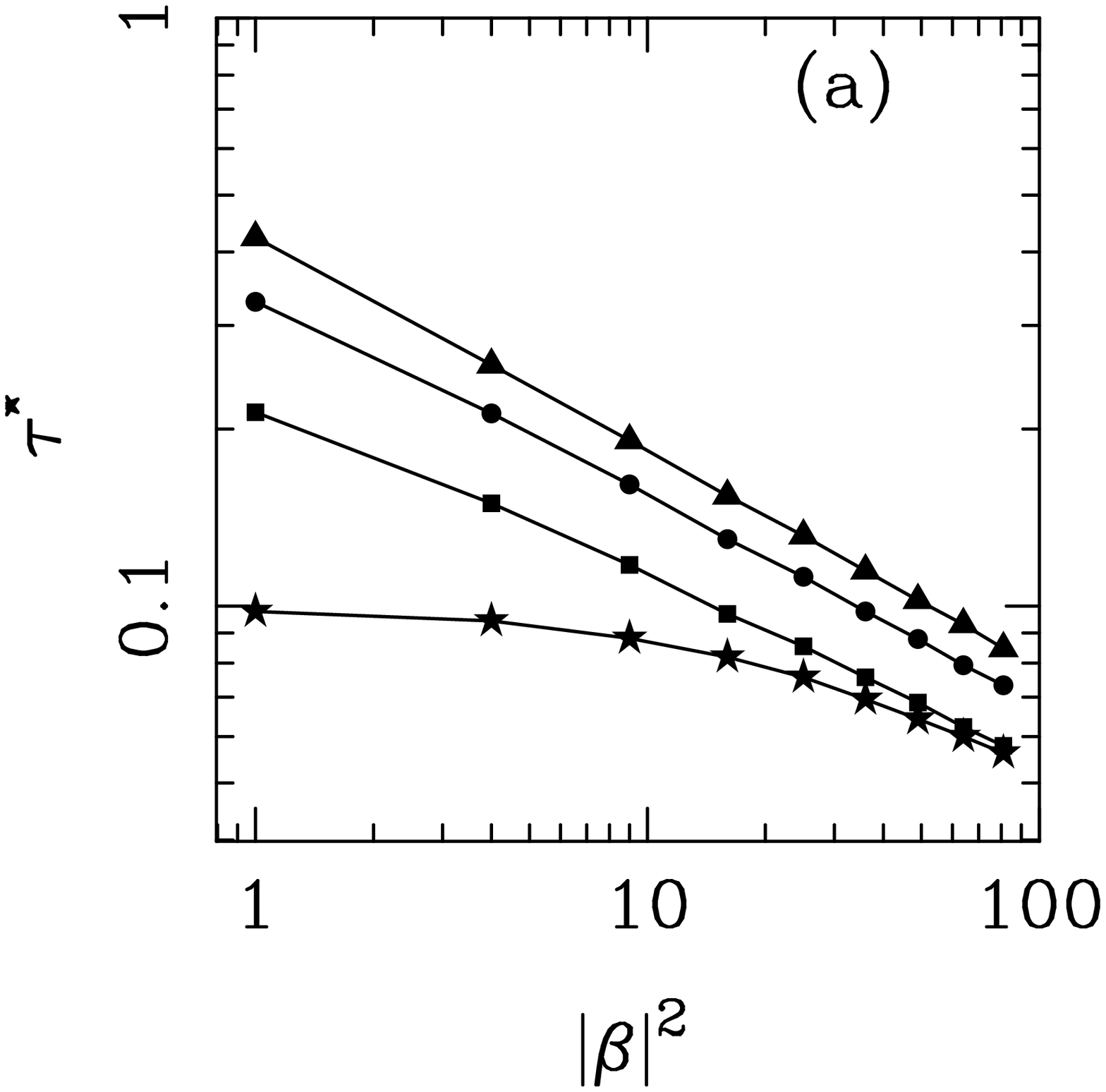}
\epsfxsize=.49\textwidth\leavevmode\epsffile{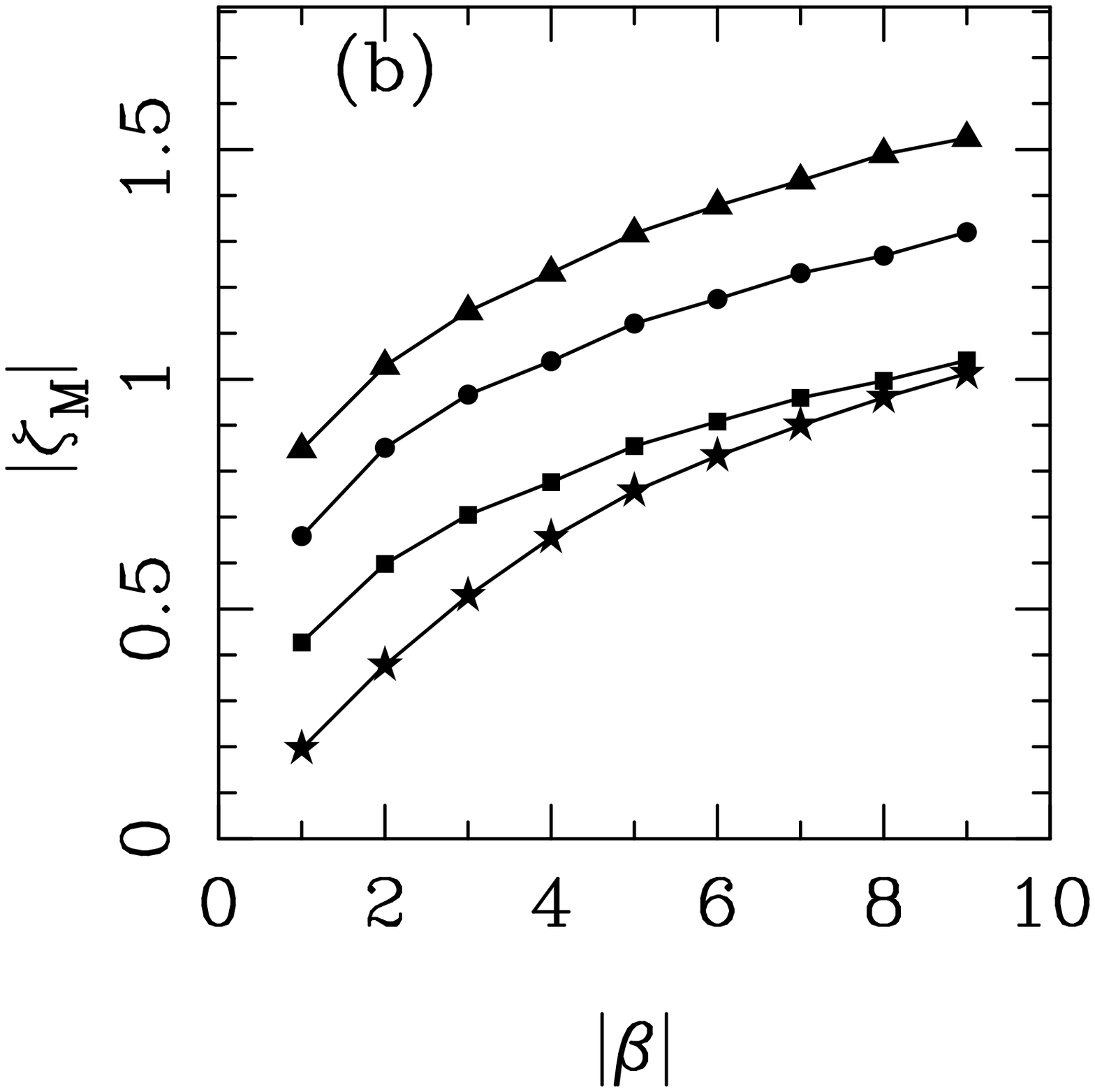}
\end{center}
\vskip .8cm
\caption{Performances of a degenerate parametric amplifier in providing 
the squeezing operator $\hat S(\zeta )$. 
In both pictures triangles refers to vacuum input, 
squares to coherent state input $\vert\alpha\equiv 1\rangle$, circles to 
number state input $\vert n\equiv 1\rangle$, and 
stars to number state input $\vert n\equiv 2\rangle$.
In (a) we report the quantity 
$\tau ^{\star}$, namely the maximum interaction time that leads to an 
output signal whose overlap with the theoretical squeezed state is 
larger than $99\%$, as a function of the pump intensity $\vert\beta\vert^2$. 
In (b) we show the maximum squeezing parameter 
$\vert\zeta_M\vert$ achievable by the degenerate parametric amplifier, 
as a function of the pump amplitude $\vert\beta\vert$.}
\label{f:psa1}
\end{figure}

\begin{figure}
\begin{center}
\epsfxsize=.49\textwidth\leavevmode\epsffile{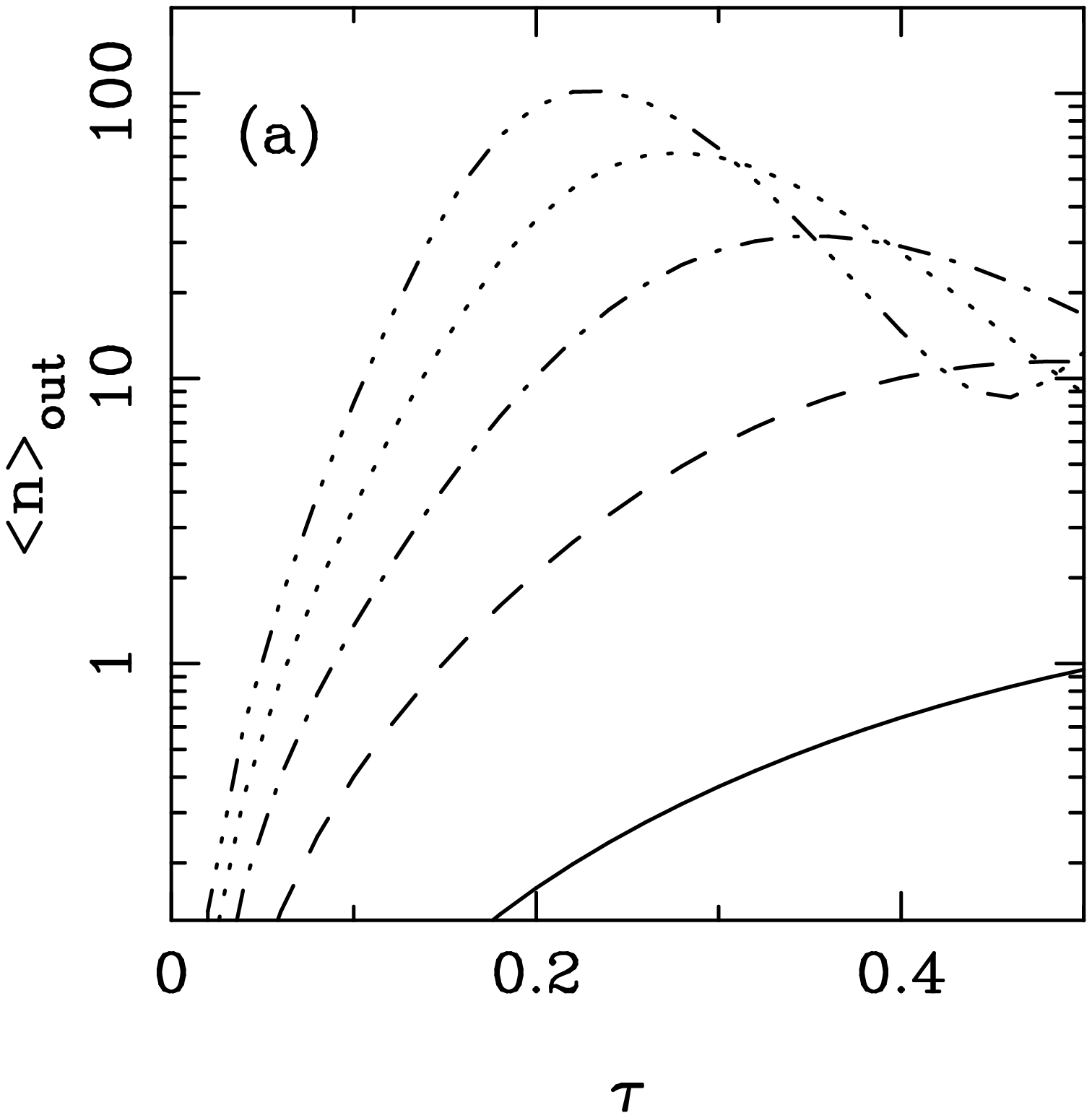}
\epsfxsize=.49\textwidth\leavevmode\epsffile{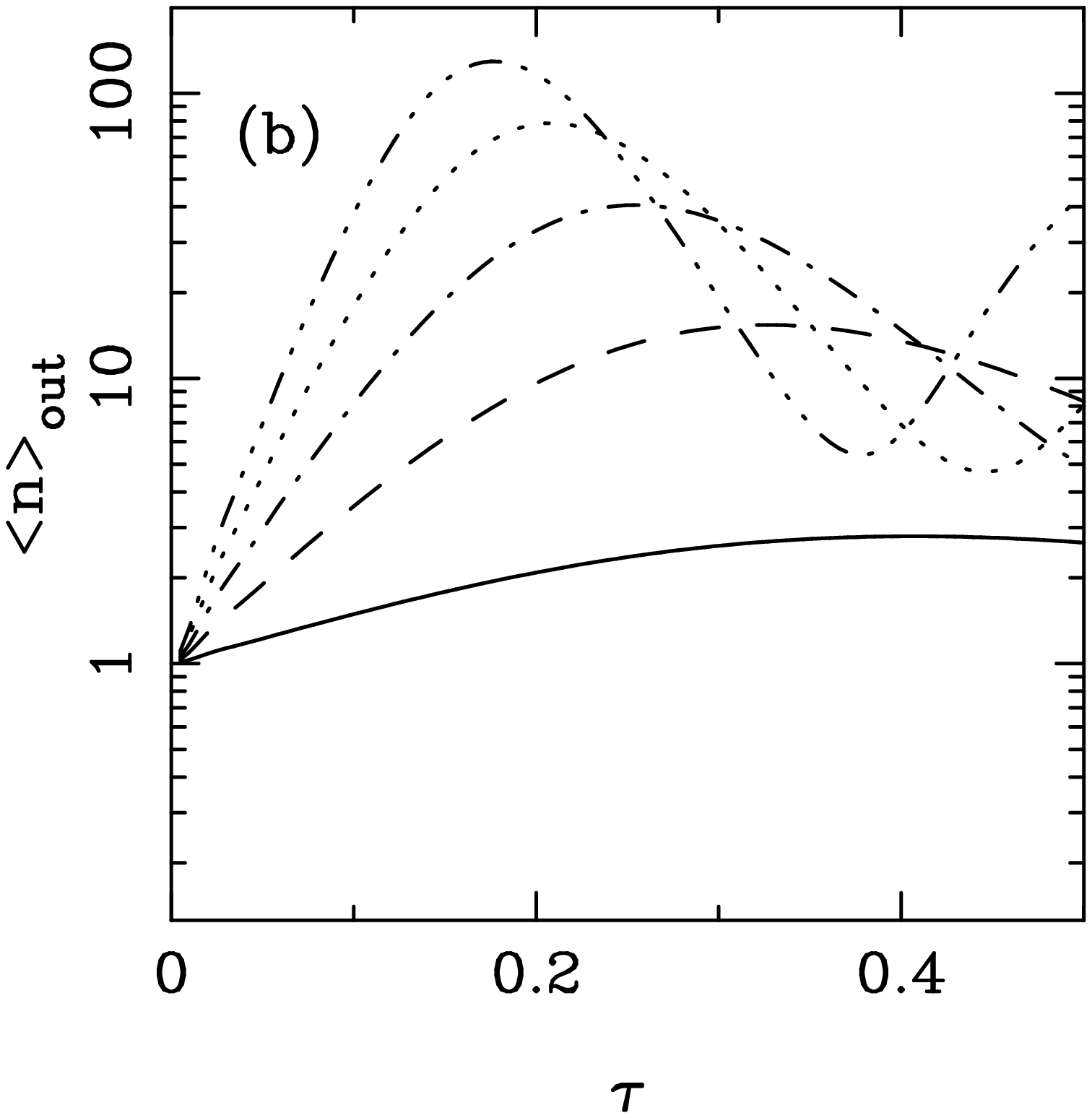}
\end{center}
\begin{center}
\epsfxsize=.49\textwidth\leavevmode\epsffile{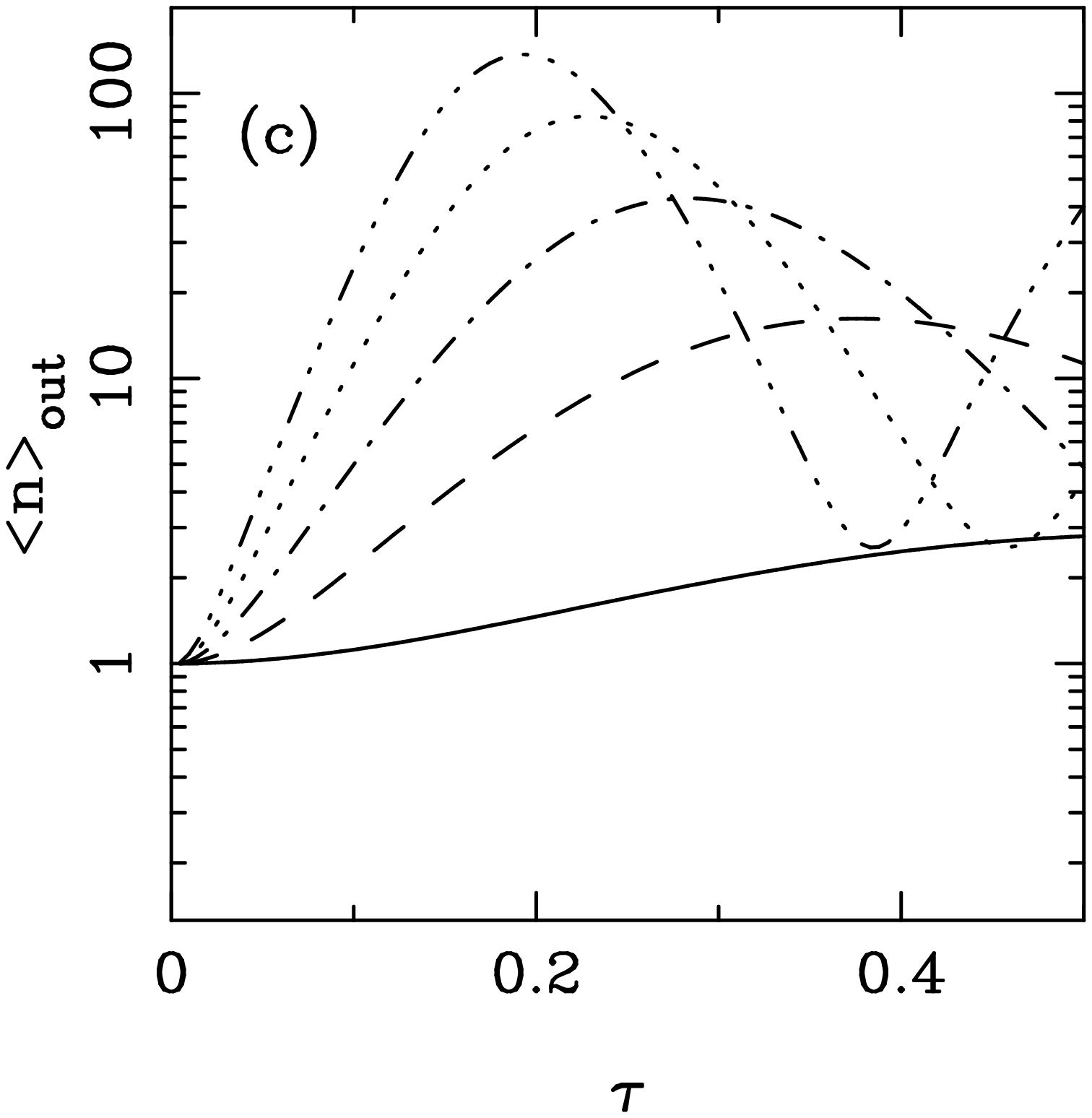}
\epsfxsize=.49\textwidth\leavevmode\epsffile{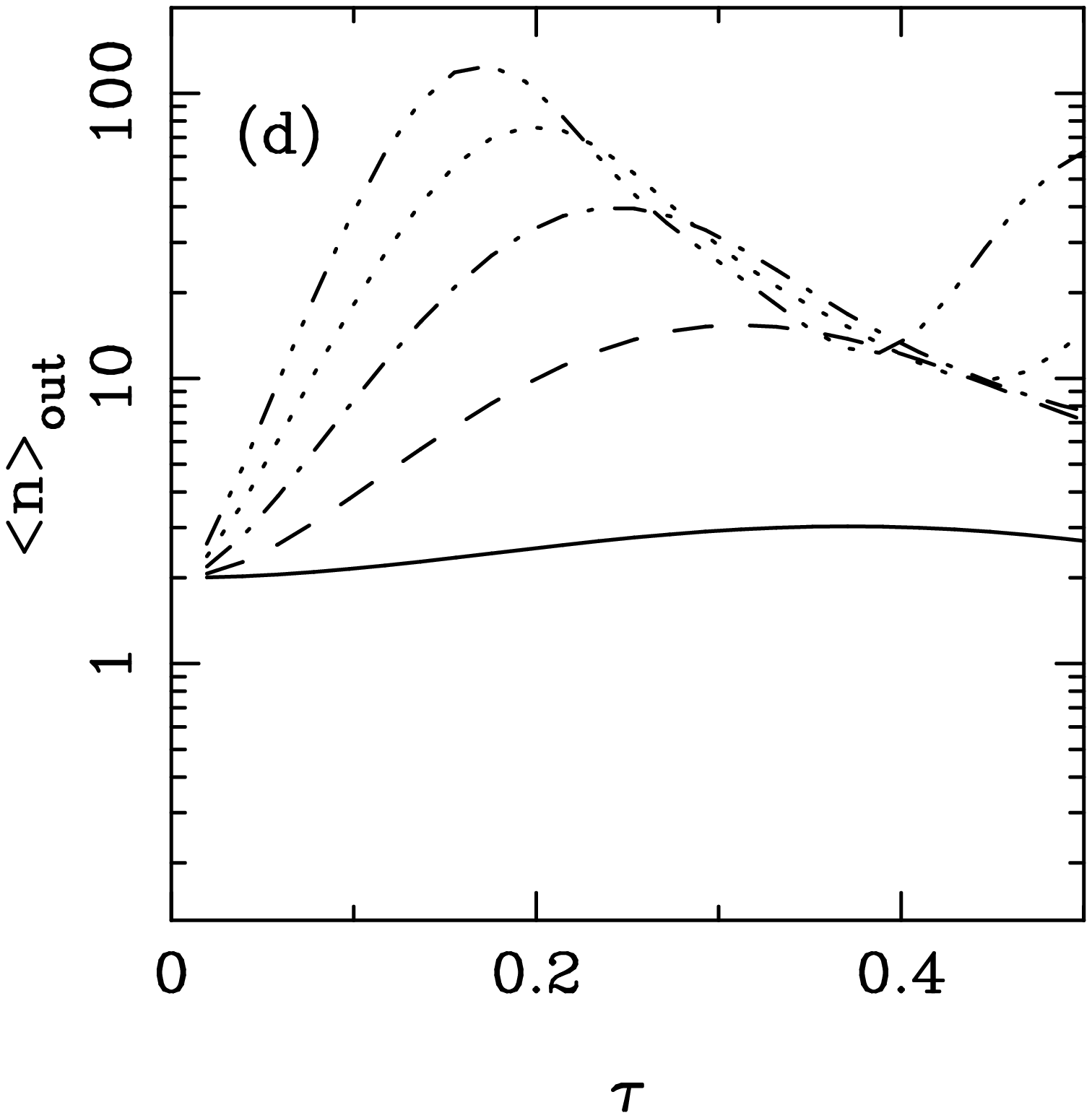}
\end{center}
\vskip .8cm
\caption{Average photon number $\langle\hat n 
\rangle_{\hbox{\scriptsize out}}$ of the signal at the output of a 
degenerate parametric amplifier, as a function of the interaction 
time $\tau$. In (a) the case of vacuum input, in (b) coherent input 
$\vert\alpha\equiv 1\rangle$, in (c) number input $\vert n\equiv 
1\rangle$, and in (d) number input $\vert n\equiv 2\rangle$. 
Different line-styles refer to different pump amplitudes: 
$\beta=9$ (dot-dot-dashed), $\beta=7$ (dotted), $\beta=5$ (dot-dashed), 
$\beta=3$ (dashed), $\beta=1$ (solid).}
\label{f:psa2}
\end{figure}

 Here we evaluate the exact evolution generated by the operator 
(\ref{udp}) through numerical diagonalization of the Hamiltonian 
(\ref{hdp}), using the method based on the constant of motion.
In this case one has
\begin{equation}
\left[\hat H _{DP},a^{\dag}a+2 c^{\dag}c\right]=0
\label{ccom}\;,
\end{equation}
and the Hilbert space ${\cal H}_a\otimes{\cal H}_c$ is decomposed in terms 
of invariant subspaces corresponding to the eigenvalues of the constant 
of motion $a^{\dag}a+2 c^{\dag}c$, namely
\begin{eqnarray}
{\cal H}_a \otimes {\cal H}_c=\oplus _{N=0}^{+\infty} 
{\cal H}_N \label{ssum}\;,
\end{eqnarray}
with
\begin{eqnarray}
{\cal H}_N=\hbox{Span}\left\{
|N-2m\rangle \otimes|m\rangle \;, m\in \left[0,\lfloor N/2\rceil
\;\right]\right\}\label{decda}\;,
\end{eqnarray}
$\lfloor \;\rceil $ denoting the integer part of $x$. 
Hence the Hamiltonian in Eq. (\ref{hdp}) rewrites in the same fashion 
as in Eq. (\ref{hn}) and the block-diagonalization is 
performed for each $\hat h_N$, with $N$ from $0$ to 
the maximum allowed value of the constant of motion. 
Similarly to Eq. (\ref{spe}), a generic two-mode state 
$\vert\psi_0\rangle$ is written as follows
\begin{eqnarray}
\vert\psi_0\rangle= \sum_{N=0}^{+\infty} \sum_{m=0}^{\lfloor N/2 \rceil }
c_{N-2m,m}\:\vert N-2m,m\rangle
\label{spe2}\;.
\end{eqnarray}                                              
The performances of a degenerate parametric amplifier in realizing the 
squeezing operator $\hat S(\zeta )= \exp [ 1/2 (\zeta a^{\dag 2} 
- \bar{\zeta} a^2)]$ of Eq. (\ref{squez}) are depicted in 
Fig. \ref{f:psa1}. In Fig. \ref{f:psa1}a we have reported the 
maximum interaction time $\tau ^{\star}$ that leads to an 
output signal whose overlap with the theoretical squeezed state is 
larger than $99\%$, as a function of the pump intensity $\vert\beta\vert^2$. 
In \ref{f:psa1}b we have shown the maximum squeezing parameter 
$\vert\zeta_M\vert$ achievable by the amplifier, 
as a function of the pump amplitude $\vert\beta\vert$. According to 
Eq. (\ref{squez}) one has $\vert\zeta_M\vert=2|\beta|\tau^{\star}$. 
In both pictures we have considered the vacuum, a coherent state and two 
different number states at the input of the amplifier. 
For the same set of input states, we have also shown in Fig. \ref{f:psa2} 
the average signal photon number as a function of the interaction time 
$\tau$, for five different values of the pump amplitude. 
\par
From Fig. \ref{f:psa1} it turns out that the requirements for 
vacuum input signal in Eq. (\ref{conddp}) are too strict. 
In particular, the two conditions in the second line are not satisfied 
for $1<|\beta | <9$ [see the line with triangles in Fig \ref{f:psa1}(a)]. 
Moreover, Fig. \ref{f:psa1} shows that one can 
realize a squeezing operator even through a weak pump with just one photon. 
\par
By definition the validity of the parametric approximation is guaranteed for 
$\tau < \tau^\star$, $\tau^\star$ depending on $|\beta |$ and on the input 
state. However, we want to provide a general criterion that can be easily 
checked experimentally. As we will show in the following, the 
undepletion of the pump is not a valid criterion. 
We argue that the relevant parameter, in order to
confirm whether the parametric approximation is justified 
or not, is the degree of coherence of the pump {\em after} the nonlinear 
interaction. These statements are supported by the following numerical 
results.
\par
Let us consider the case of a number input state $|n=1\rangle$ with pump 
amplitude $|\beta |=9$. 
From Fig. \ref{f:psa1} one can extract the maximum interaction time 
$\tau ^{\star}\simeq 0.073$ for the validity of the parametric approximation, 
and the corresponding maximum squeezing parameter $|\zeta_M|\simeq 1.314$. 
The average photon number of the output state can be drawn from 
Fig. \ref{f:psa2}c as $\langle n\rangle _{\hbox{\scriptsize{out}}}
\simeq 9.68$, corresponding to a pump depletion of about $5.4\%$. 
One might consider such a small depletion as the sign of the goodness of the 
parametric approximation. On the other hand, from Fig. \ref{f:psa2}c one 
recognizes the region $0.33 \lesssim\tau \lesssim 0.44$, in which the 
output signal is even less excited than that in the above example, 
and consequently the pump is less depleted. Nevertheless the parametric 
approximation does not hold in such range of interaction time, 
since $\tau$ is larger than $\tau ^{\star}$. 
Let us now consider the Fano factor $F=\langle\Delta \hat n^2\rangle/
\langle\hat n\rangle$ of the pump at the output. 
One finds that in the region $0.33 \lesssim\tau \lesssim 0.44$ 
the Fano factor is always larger than $F=1.13$, whereas for
$\tau < \tau^\star$ it never exceeds $F=1.10$. 
\par
\begin{figure}
\begin{center}
\epsfxsize=.49\textwidth\leavevmode\epsffile{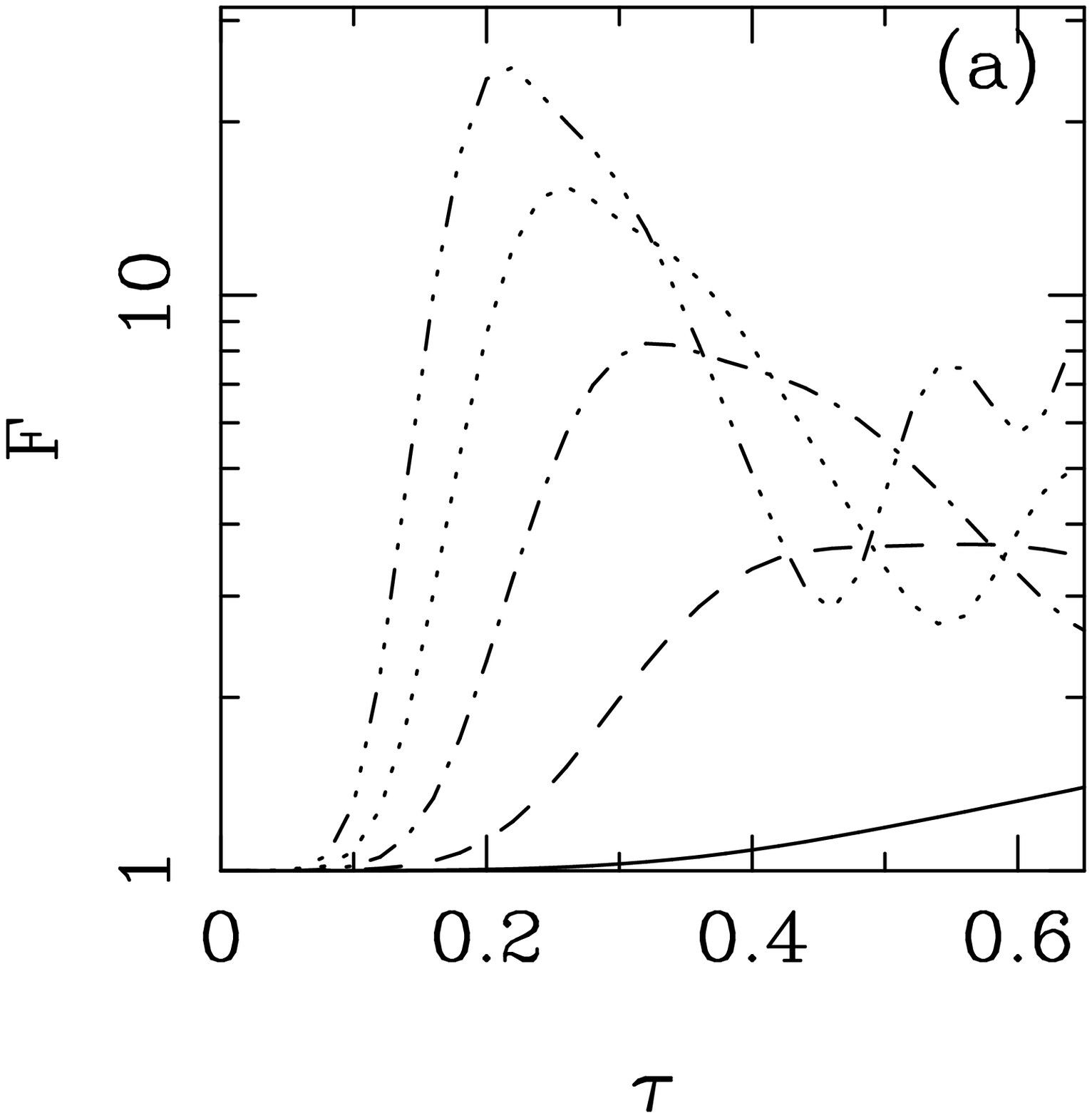}
\epsfxsize=.49\textwidth\leavevmode\epsffile{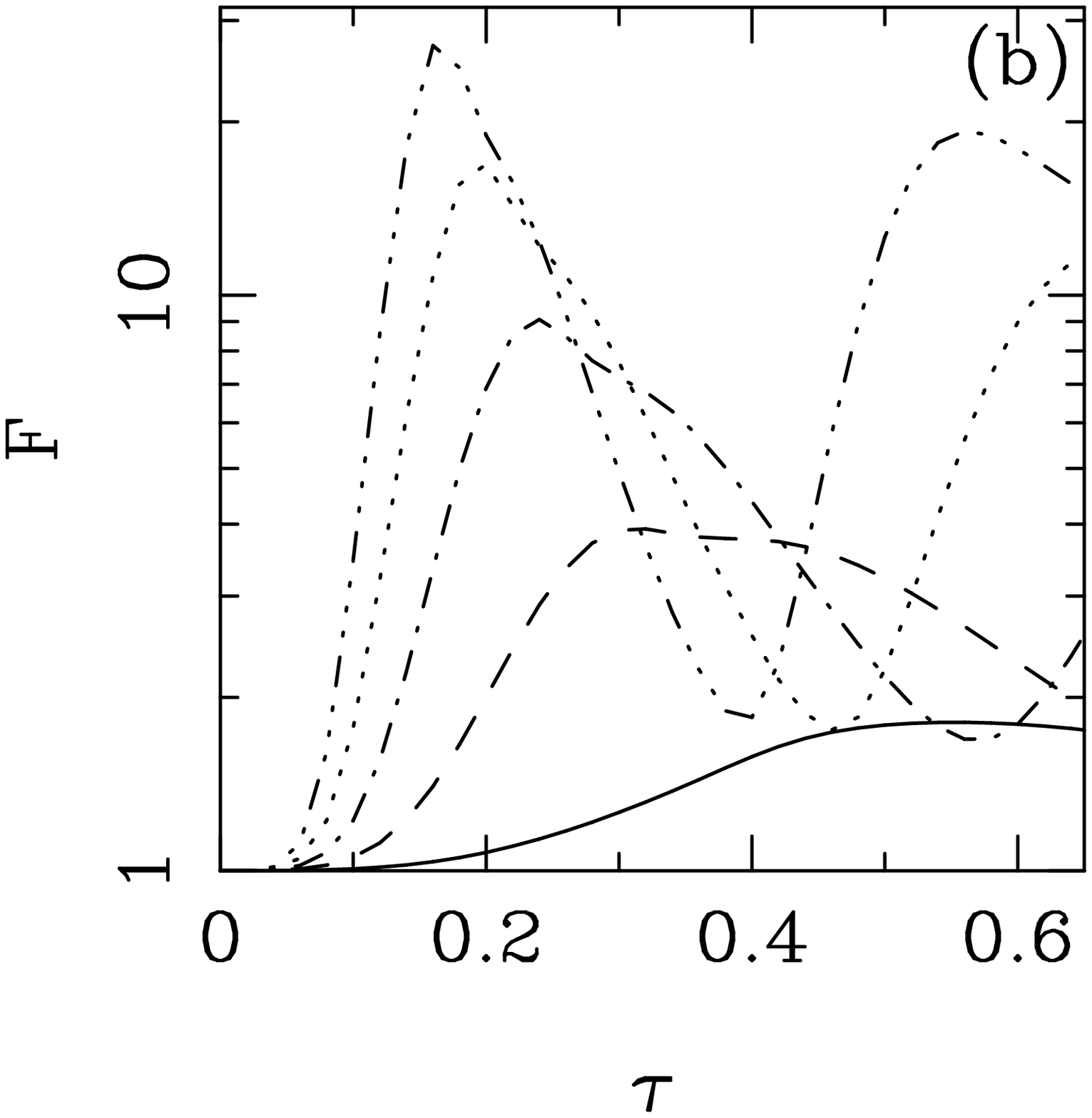}
\end{center}
\vskip .8cm
\caption{Fano factor $F$ of the pump at the output of a degenerate parametric 
amplifier, as a function of the interaction time $\tau$. 
In (a) the case of vacuum input; in (b) of coherent input 
$\vert\alpha\equiv 1\rangle$. Different line-styles refer to different 
pump amplitude: $\beta=9$ (dot-dot-dashed), $\beta=7$ (dotted), 
$\beta=5$ (dot-dashed), $\beta=3$ (dashed), $\beta=1$ (solid).}
\label{f:psa5}
\end{figure}

More generally, in all situations in which the parametric 
approximation is satisfied we found that the Fano factor of the pump 
at the output never exceeds $F=1.10$. This holds also when the pump is weak 
($|\beta |^2=1 \div 10$). Indeed, in this case the depletion of the pump 
can be strong, nevertheless the parametric approximation does not break down. 
In Fig. \ref{f:psa5} we show the Fano factor of the output pump as a function 
of the interaction time $\tau$, for different values of the pump amplitude. 
Plots refer to vacuum input and to coherent state input 
$\vert \alpha\equiv 1\rangle$: similar plots can be obtained for other input 
states.
\par 
As the condition of pump undepletion does not guarantee the validity of the 
parametric approximation, so pump depletion by itself does not sign its 
failure: rather we have to consider the Fano factor of the pump. 
In order to stress this point, let us consider the extreme case of a pump 
with only one photon, and the input signal in the vacuum. The exact numerical 
solution indicates that the parametric approximation holds for interaction 
time up to $\tau ^{\star }\simeq 0.42$, the squeezing parameter and the 
output signal photon number increasing up to $|\zeta_M|\simeq 0.84$ and 
$\langle n\rangle _{\hbox{\scriptsize{out}}}\simeq 0.74$, respectively 
[see Figs. \ref{f:psa1} and \ref{f:psa2}a]. 
Correspondingly, the pump depletion grows up to $37\%$ at $\tau=\tau^{\star}$. 
In spite of the strong depletion, the pump 
preserves a good degree of coherence: the Fano factor achieves at most the 
value $F=1.10$ at $\tau=\tau^{\star}$. 
\par
In summary, the validity regime $\tau < \tau^\star$ for the parametric 
approximation does not identify with the 
condition of pump undepletion, rather it corresponds to a Fano factor not 
exceeding the initial coherent level more than $10\%$. 
\par
\begin{figure}
\begin{center}
\epsfxsize=.49\textwidth\leavevmode\epsffile{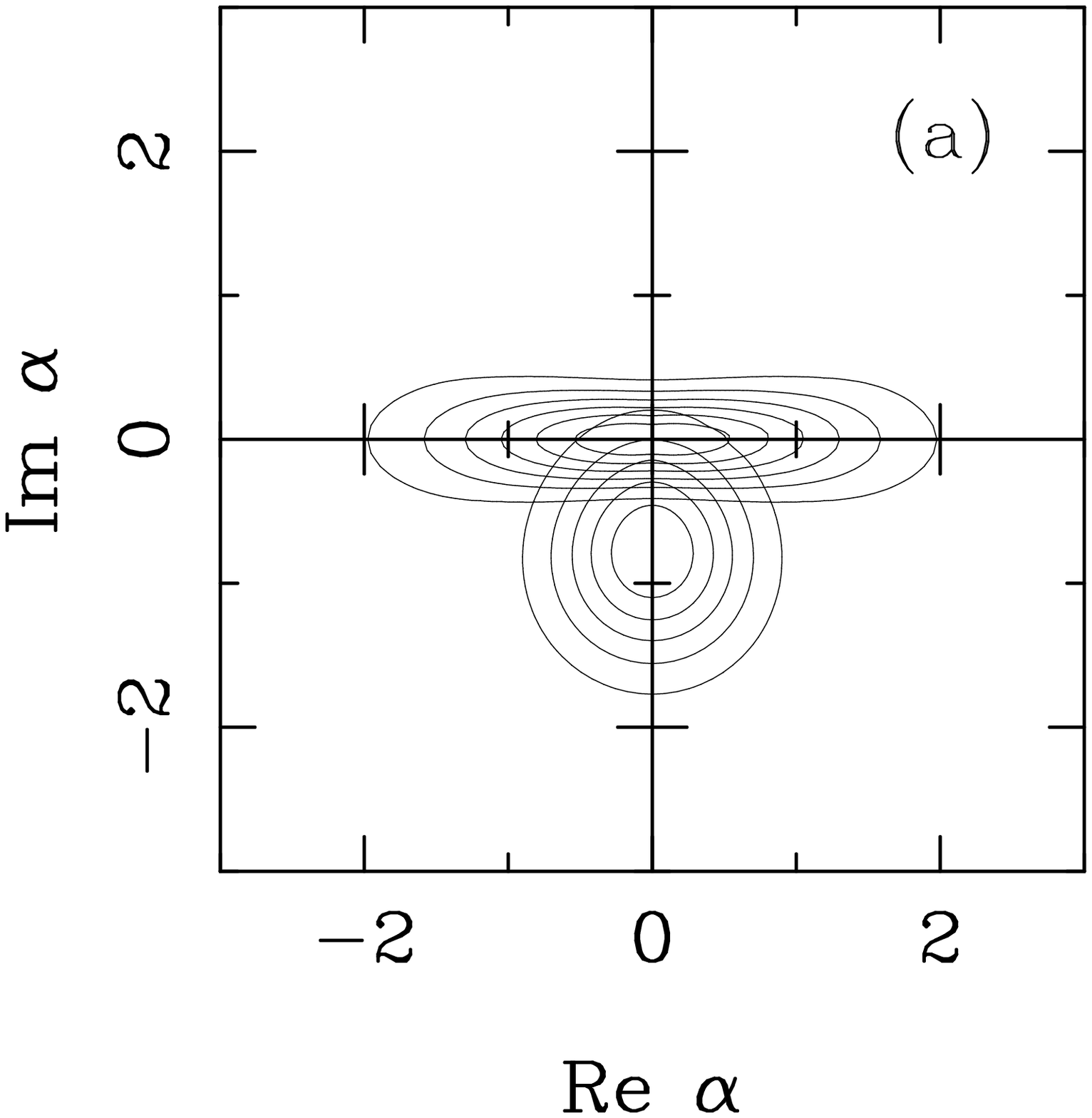}
\epsfxsize=.49\textwidth\leavevmode\epsffile{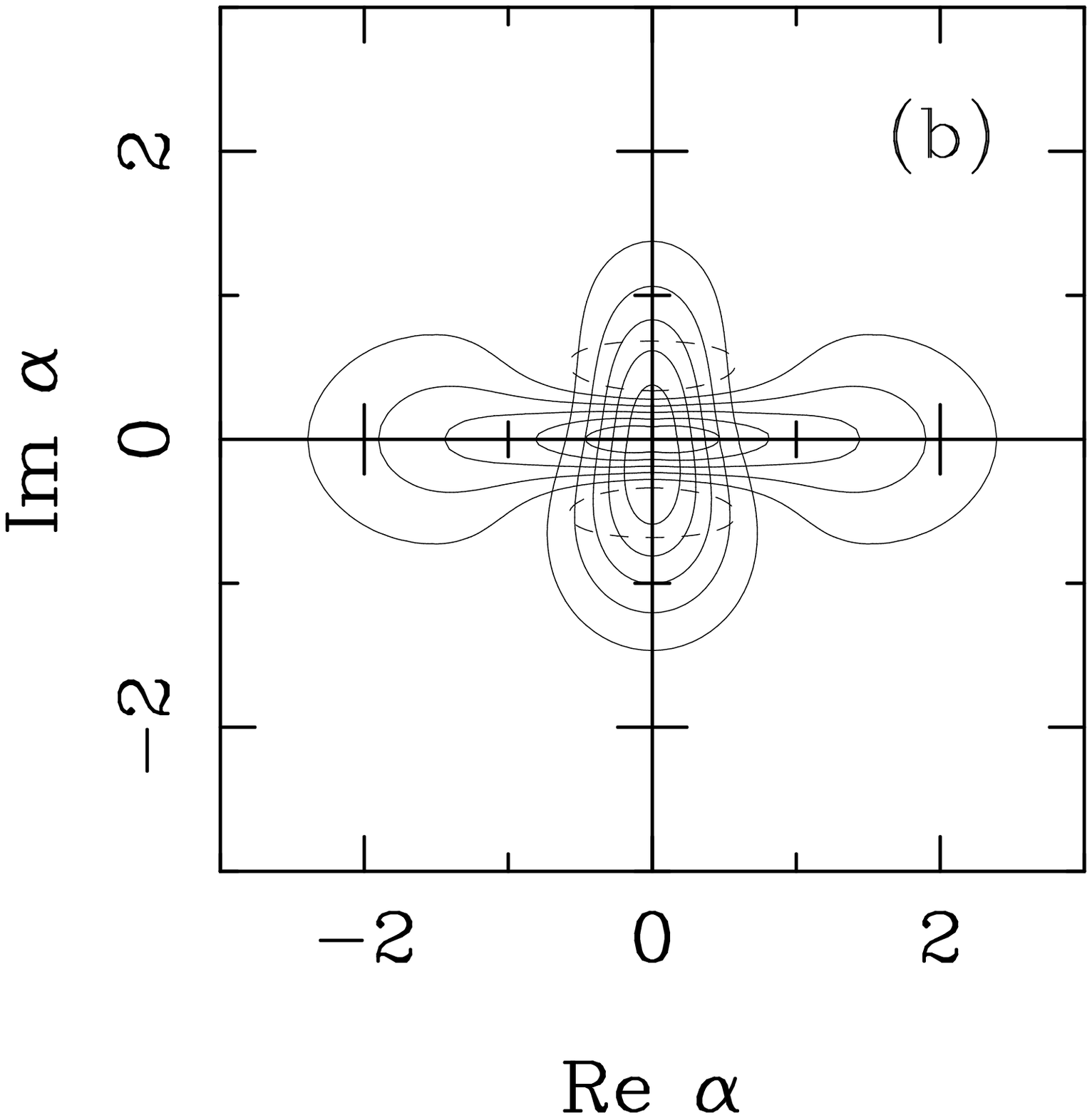}
\end{center}
\vskip .8cm
\caption{Contour plot Wigner functions of both the signal and the pump 
modes at the output 
of a degenerate parametric amplifier. The input signal is the 
vacuum, whereas the pump is initially in a coherent state $\beta=-i$. 
The time interaction is equal to $\tau=\tau^\star=0.42$ in (a) and to 
$\tau=2\tau^\star=0.84$ in (b).}
\label{f:psa3}
\end{figure}

\begin{figure}
\begin{center}
\epsfxsize=.49\textwidth\leavevmode\epsffile{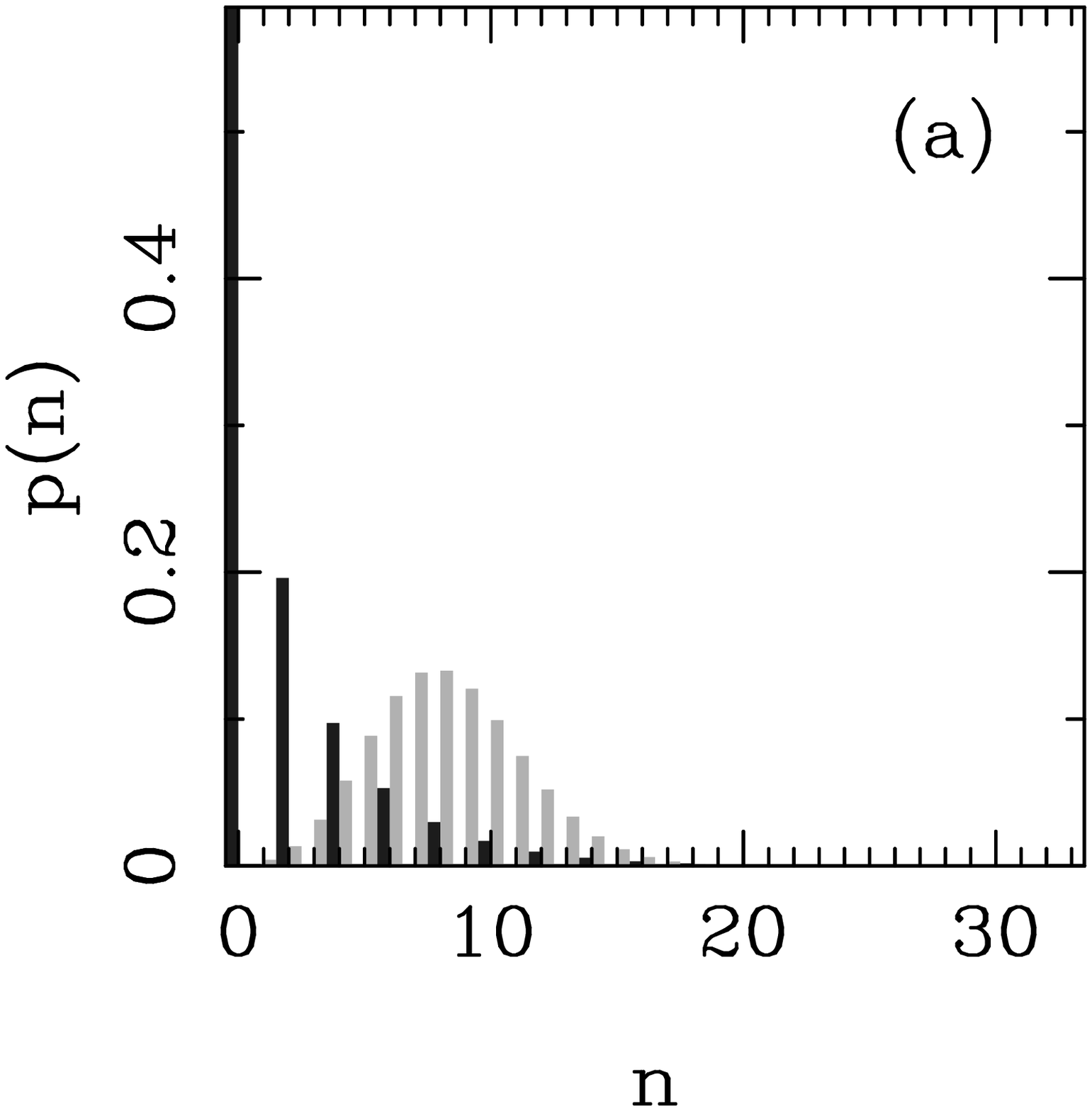}
\epsfxsize=.49\textwidth\leavevmode\epsffile{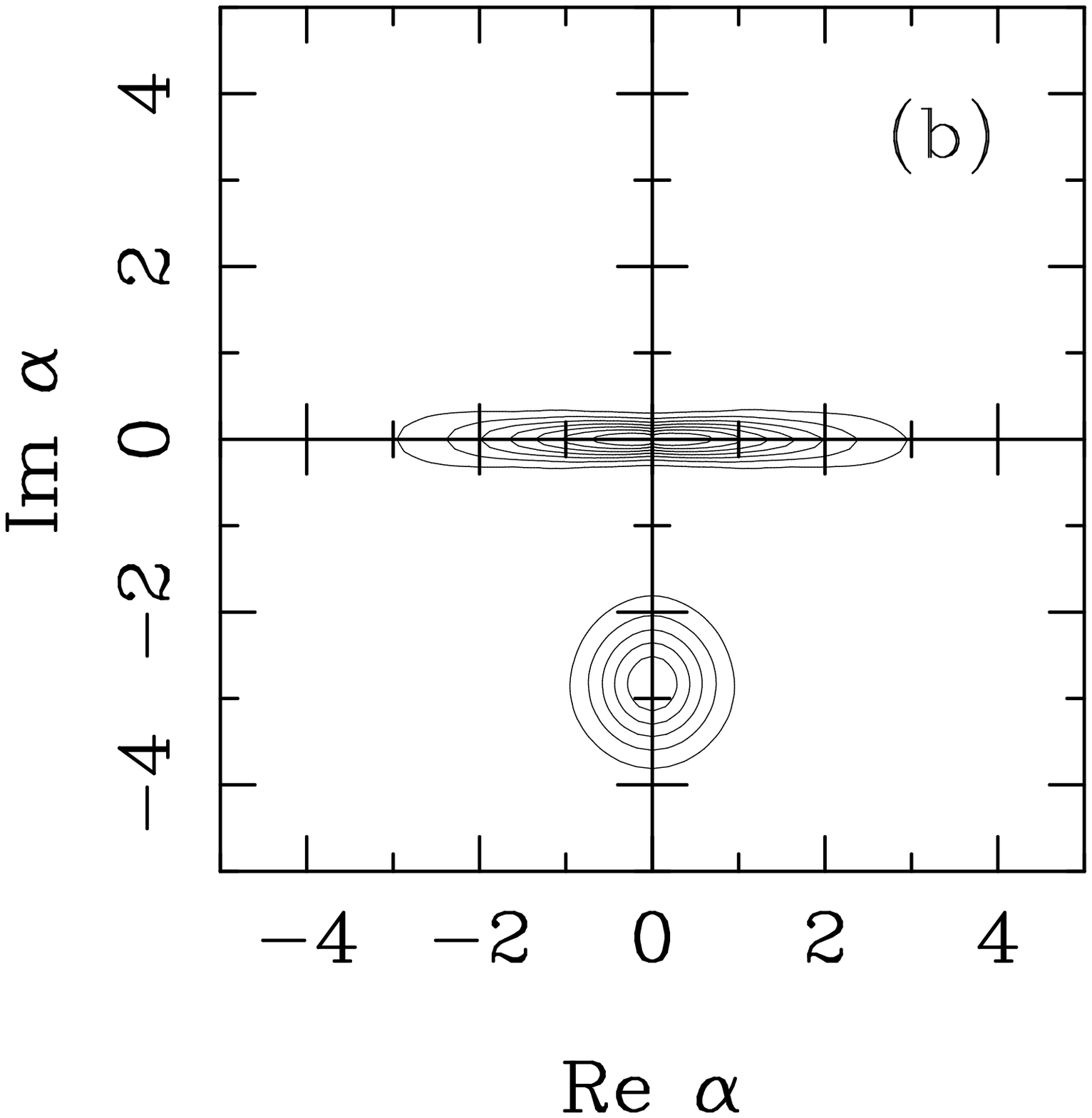}
\end{center}
\begin{center}
\epsfxsize=.49\textwidth\leavevmode\epsffile{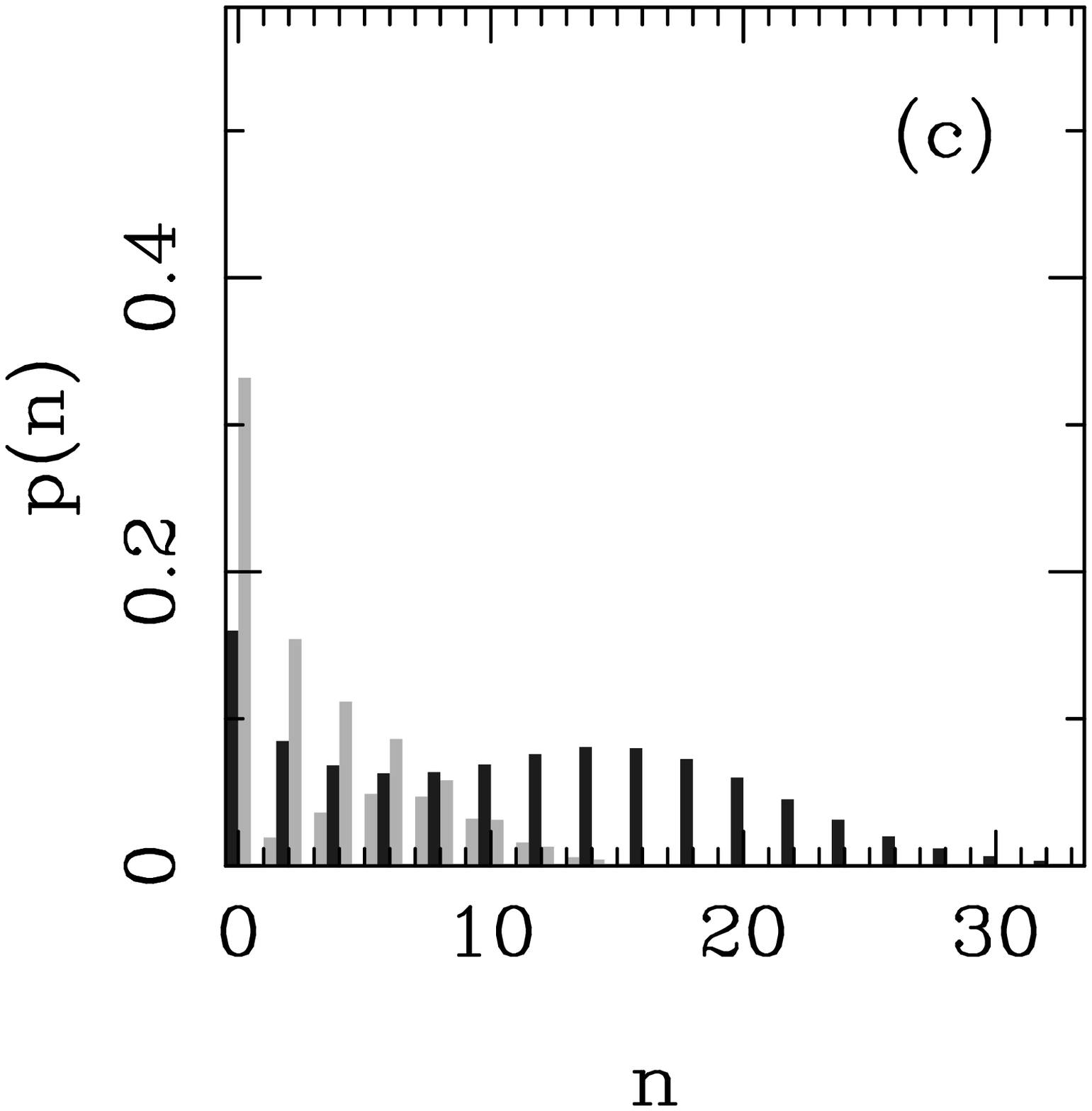}
\epsfxsize=.49\textwidth\leavevmode\epsffile{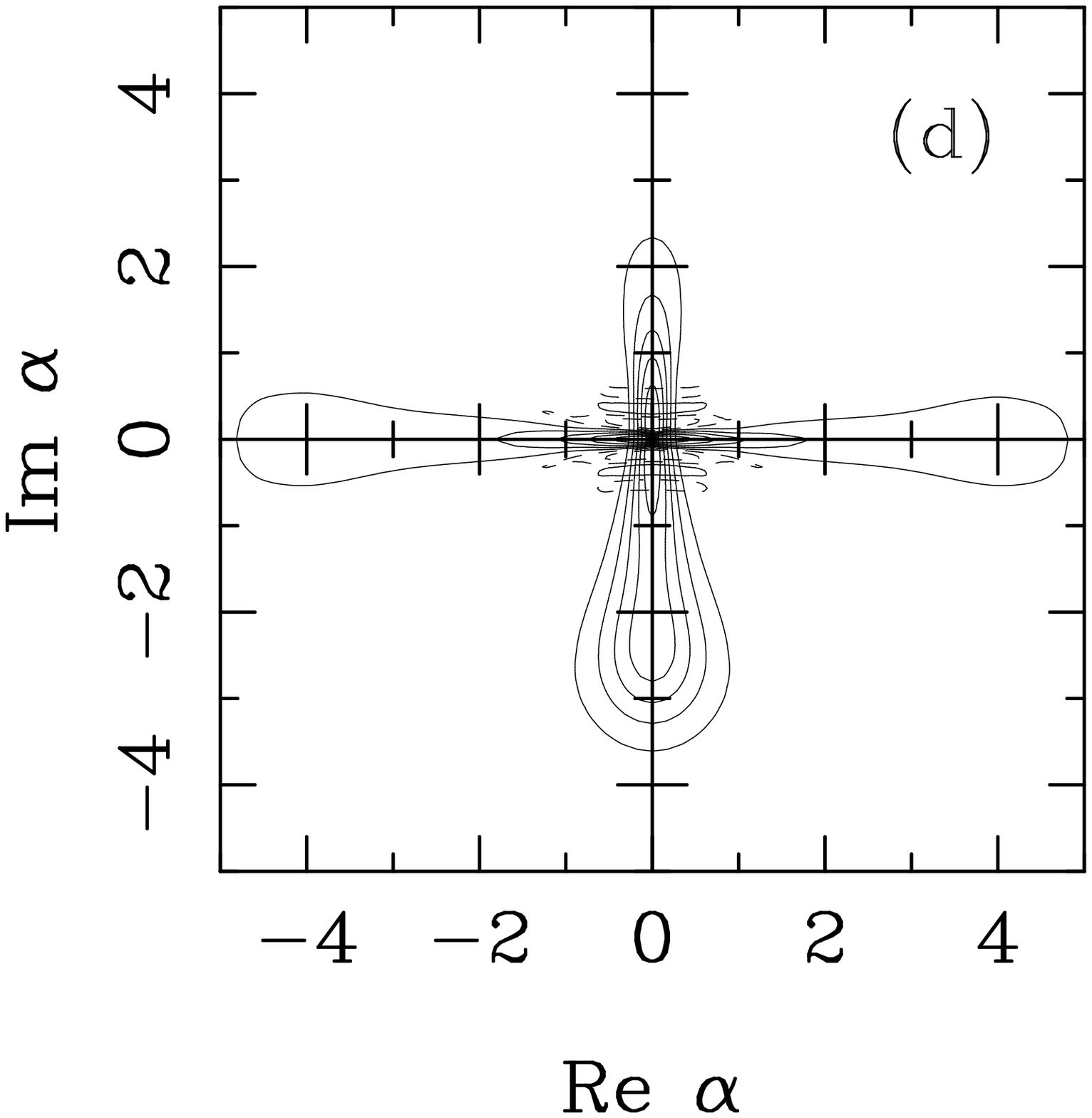}
\end{center}
\vskip .8cm
\caption{Photon number probability and contour plot Wigner function
for both the signal and the pump mode at the output of a degenerate
parametric amplifier.  The plots refer to a situation in which the
signal mode is initially in the vacuum and the pump mode is excited to
a coherent state with amplitude $\beta=-3i$. In (a) and (b) the
interaction time is equal to $\tau\equiv\tau^\star=0.19$, whereas (c)
and (d) refer to an interaction time $\tau= 0.43 > 2 \tau^\star$. In
the first case the parametric approximation well describes the real
interaction, which produces a squeezed vacuum state with squeezing
parameter $r=1.146$ corresponding to about $2$ squeezing photons. On
the other hand, parametric approximation does not hold in the second
case. Notice that the break-down of parametric approximation is 
connected with the appearance of negative values in the Wigner
function, which is a signature of quantum interference in the phase
space.}
\label{f:psa4}
\end{figure}

 
What happens beyond the parametric regime? For interaction time larger 
than $\tau^{\star}$ new quantum effects arise at the output. In Fig. 
\ref{f:psa3} we show the Wigner functions of both
the signal and the pump modes at the output of the amplifier for 
$\tau=\tau^\star$ and $\tau=2\tau^\star$, with vacuum input and weak pump 
$|\beta |=1$. In Fig. \ref{f:psa4} the case of a stronger pump is given. 
As $\tau$ increases, 
the pump first empties, then it starts refilling, preferably for even photon 
numbers, leading to oscillations in the photon number distribution. 
Remarkably, the corresponding Wigner functions of the pump and the signal 
exhibit interference in the phase space \cite{schl}, 
the signal resembling a Schr\"odinger-cat.
\section{Two-mode squeezer}
The nondegenerate parametric amplifier involves three 
different modes of the radiation field---say the signal $a$, the idler $b$ and 
the pump $c$---which are coupled by a $\chi^{(2)}$ nonlinear medium. 
The relation between the frequencies of the field modes is given by 
$\omega _c=\omega _a+\omega _b$. 
The Hamiltonian of the amplifier under phase-matched conditions 
can be written in the rotating wave 
approximation as follows
\begin{eqnarray}
\hat H_{NP}=\kappa \left(abc^{\dag }+a^{\dag }b^{\dag } c\right)\label{hnp}\;,
\end{eqnarray}
with $\kappa \propto \chi ^{(2)}$. The Hamiltonian in Eq. (\ref{hnp}) 
describes also the case in which the frequencies pertaining modes $a$ and $b$ 
are the same, provided that the respective wave vectors and/or polarizations 
are different. 
The dynamics induced by the Hamiltonian (\ref{hnp}) leads to a considerably 
rich variety of phenomena, such as generation of strongly correlated photon 
pairs by parametric downconversion 
\cite{shih,ou,rarity}, phase insensitive amplification 
\cite{rev,mol}, generation of heterodyne eigenstates that are 
suitable for optimal phase detection \cite{sac}, 
polarization entanglement \cite{dema} and realization of 
Bell states \cite{rarity,dema,kly,horne}. 
The unitary evolution operator in the interaction picture reads 
\begin{eqnarray}
\hat U _{NP}=\exp\left[-i \tau\left(abc^{\dag }+a^{\dag }b^{\dag } c\right)
\right]
\label{unp}\;,
\end{eqnarray}
where $\tau$ represents a rescaled interaction time. 
The parametric approximation replaces in Eq. (\ref{unp}) 
the pump mode $c$ with the complex amplitude $\beta$ of the 
corresponding coherent state, thus achieving  the two-mode squeezing operator
\begin{eqnarray}
\hat S_2(-i\tau\beta)\equiv 
\exp\left[-i\tau\left(\beta a^{\dag }b^{\dag }+\bar\beta 
ab \right) \right]\label{stw}\;.
\end{eqnarray}
The two-mode squeezing operator yields a suppression of the quantum 
fluctuations in one quadrature of the sum and difference of modes 
$a\pm b$ \cite{caves}. When applied to vacuum input, 
the unitary operator in Eq. (\ref{stw}) generates the so-called twin-beam 
\begin{eqnarray}
\hat S_2(\chi)|0,0\rangle= (1-|\lambda|^2)^{1/2}\sum_{n=0}^{\infty}
\lambda ^n |n,n\rangle \label{twb}\;,
\end{eqnarray}
where $\lambda =\arg(\chi)\tanh |\chi|$. The expression in Eq. (\ref{twb}) 
can be easily derived by factorizing  the $\hat S_2$ operator through 
the decomposition formulas for the SU(1,1) Lie algebra \cite{rev,tru,su}. 
\par The dynamics of the nondegenerate parametric amplifier admits two 
independent constants of motion. 
We choose them as follows 
\begin{eqnarray}
\hat N = \frac{1}{2}\left[ a^{\dag} a+ b^{\dag} b + 
2 c^{\dag} c \right]\:,  \quad\quad
\hat K =a^{\dag} a + c^{\dag} c 
\label{kost}\:.
\end{eqnarray}
Correspondingly, we decompose the Hilbert space 
${\cal H}_a \otimes {\cal H}_b \otimes {\cal H}_c$ in the direct sum
\begin{eqnarray}
{\cal H}_a \otimes {\cal H}_b \otimes {\cal H}_c =\oplus_{N=0}^{\infty} 
\oplus_{K=0}^{N} {\cal H}_{NK}
\label{decpia}\;, 
\end{eqnarray}
where the invariant subspaces ${\cal H}_{NK}$ are given by
\begin{eqnarray}
{\cal H}_{NK}&=&\hbox{Span}\left\{|K-m\rangle \otimes \vert N-K-m\rangle 
\otimes \vert m\rangle\;,\right.\nonumber \\ 
&&\left.m\in [0,min(K,N-K)]\right\}
\label{depia2}\;. 
\end{eqnarray}
The Hamiltonian $\hat H_{NP}$ and a generic three-mode state 
$\vert\psi_0\rangle$ will be consistently written as follows
\begin{eqnarray}
\hat H_{NP}= \sum_{N=0}^{+\infty}\sum_{K=0}^{N}  \hat h_{NK}
\label{decham}\;, 
\end{eqnarray}
\begin{eqnarray}
\vert\psi_0\rangle=\sum_{N=0}^{+\infty}\sum_{K=0}^{N}\sum_{m=0}^
{\min(K,N-K)} c_{K-m,N-K-m,m}\:\vert K-m,N-K-m,m\rangle
\label{spe3}\;.
\end{eqnarray}
To compute the exact dynamical evolution, one then diagonalizes 
each block $\hat h_{NK}$ in Eq. (\ref{decham}) up to a fixed maximum value of 
$N$ and makes the input state evolve in the representation of Eq. 
(\ref{spe3}). 
\par
\begin{figure}
\begin{center}
\epsfxsize=.49\textwidth\leavevmode\epsffile{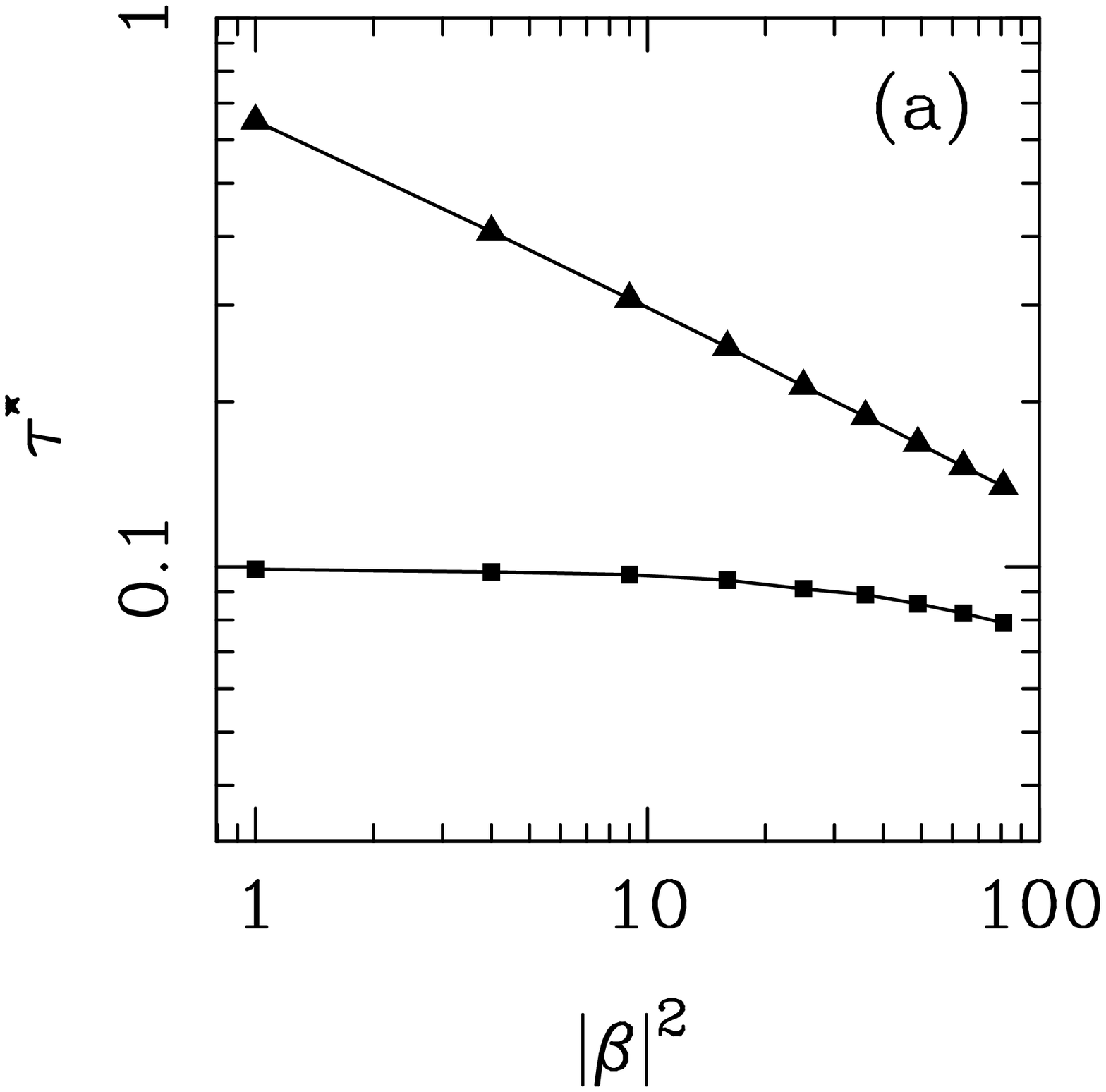}
\epsfxsize=.49\textwidth\leavevmode\epsffile{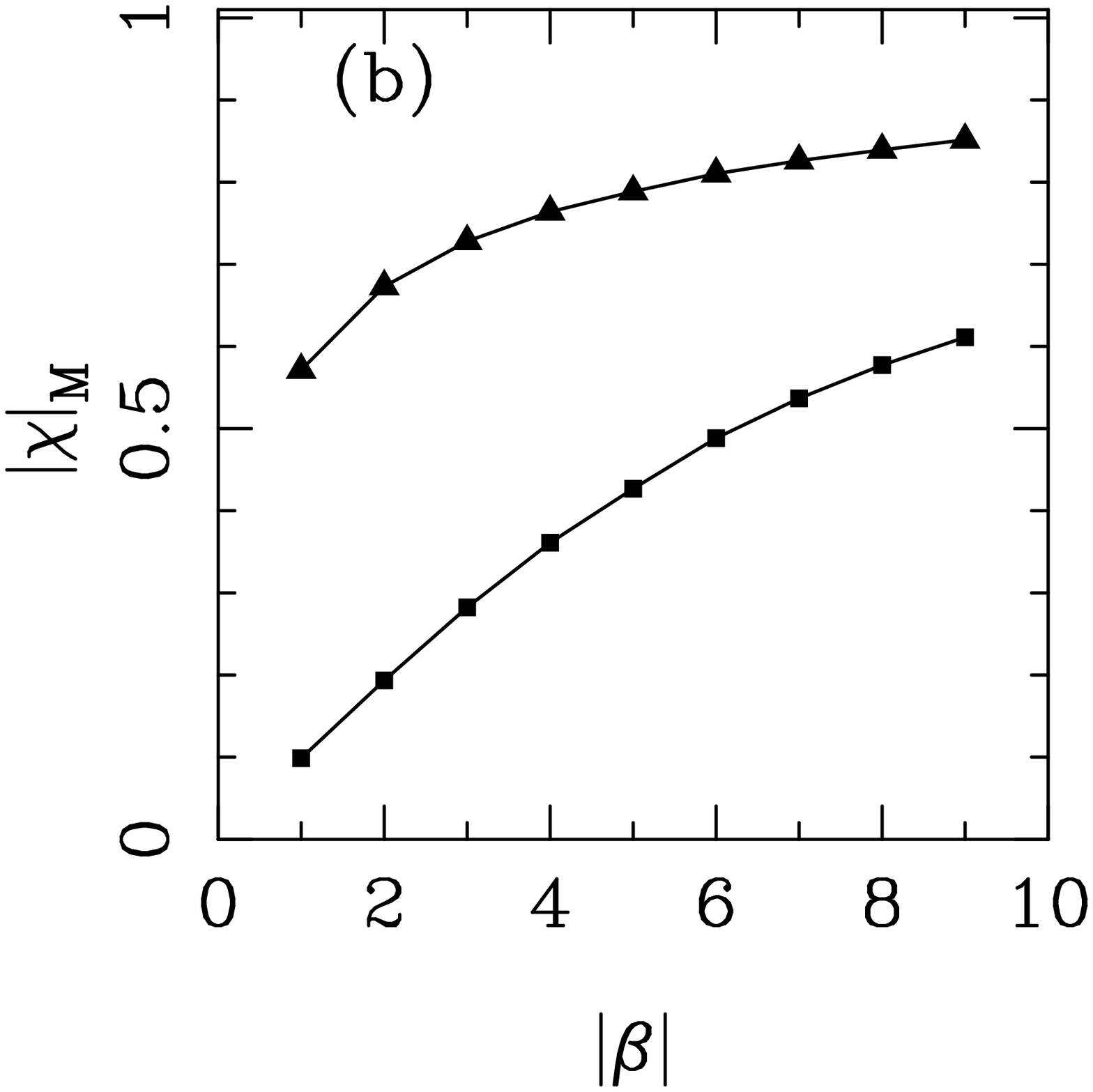}
\end{center}
\vskip .8cm
\caption{Performances of a nondegenerate parametric amplifier in achieving 
the two-mode squeezing operator. 
In both pictures triangles refers to vacuum input and 
circles to photon number state $\vert 1,1\rangle$ 
input. In (a) we report the quantity 
$\tau^{\star}$, 
namely the maximum interaction time that leads to an output signal whose 
overlap with the theoretical state is larger than $99\%$, as a 
function of the pump intensity $\vert\beta\vert^2$. 
In (b) we show the corresponding maximum two-mode squeezing 
parameter $\vert\chi_M\vert$ 
achievable by the nondegenerate parametric amplifier, as 
a function of the pump amplitude $\vert\beta\vert$.}
\label{f:pia1}
\end{figure}

\begin{figure}
\begin{center}
\epsfxsize=.49\textwidth\leavevmode\epsffile{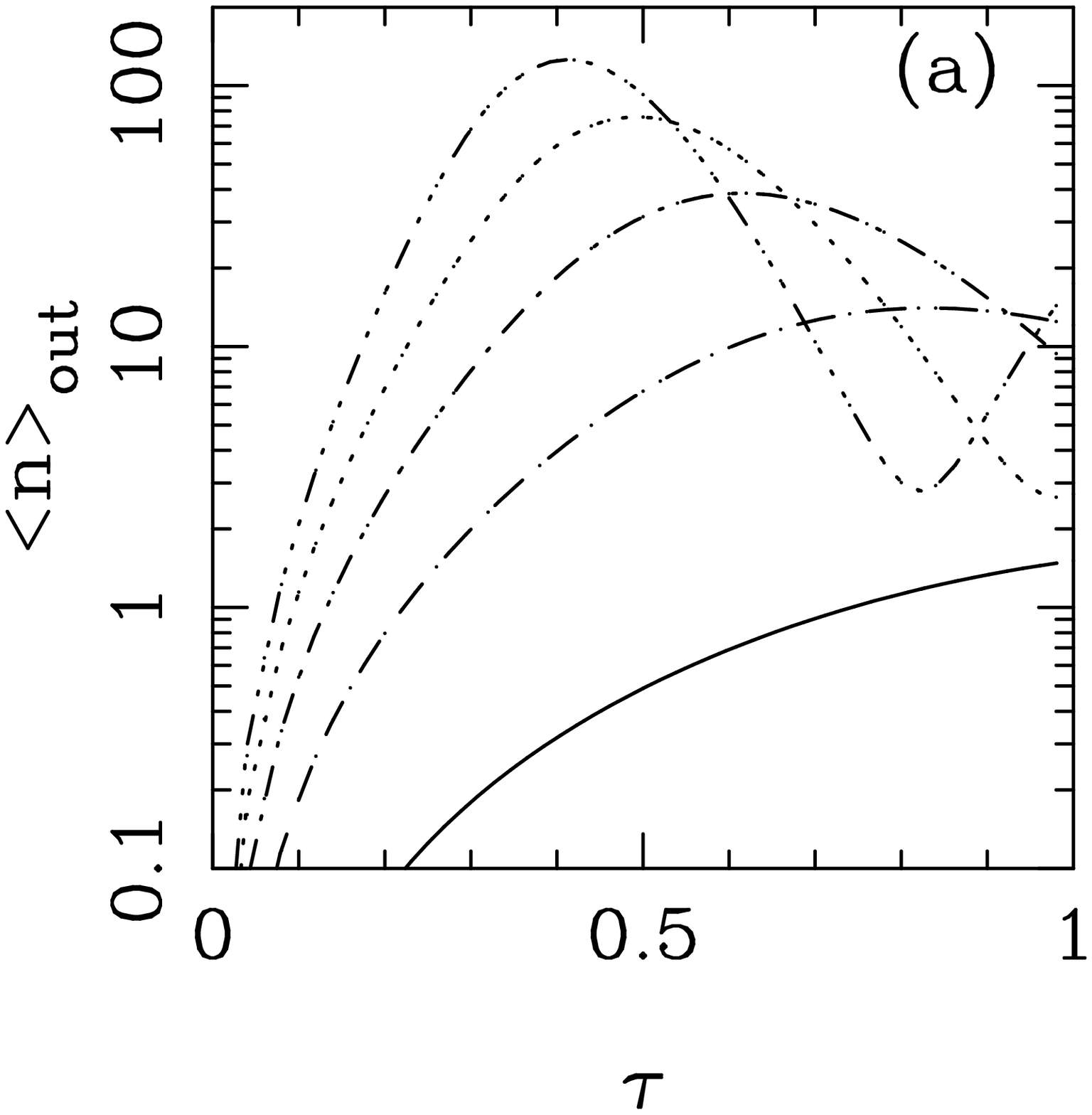}
\epsfxsize=.49\textwidth\leavevmode\epsffile{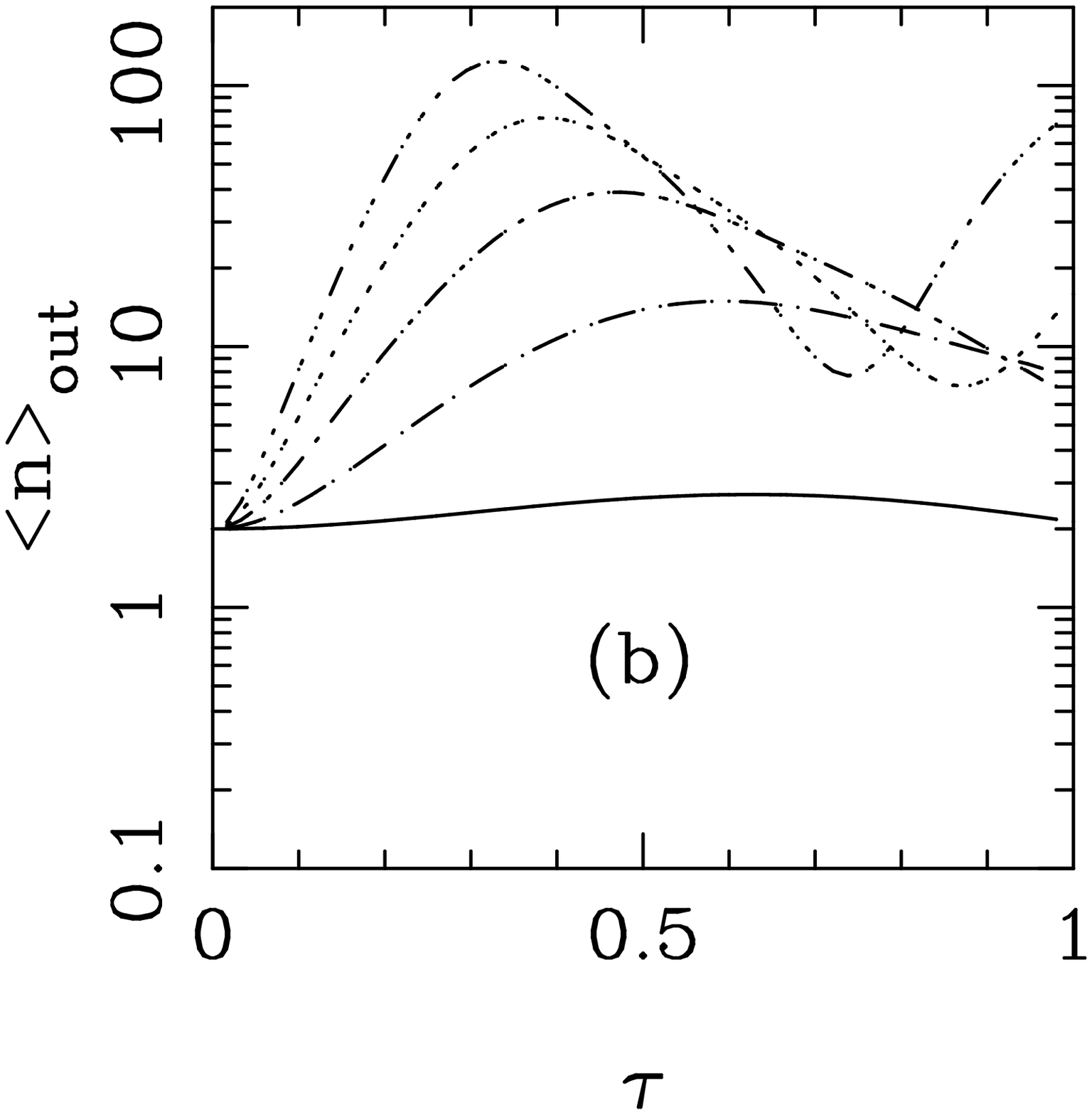}
\end{center}
\vskip .8cm
\caption{Average photon number 
$\langle\hat n\rangle_{\hbox{\scriptsize out}}$ 
of the signal at the output of 
a nondegenerate parametric amplifier, as a function of 
the interaction time $\tau$.
In (a) the case of vacuum input, and in (b) the case of photon number 
input $\vert 1,1 \rangle$. 
Different line-styles refer to different pump amplitudes: 
$\beta=9$ (dot-dot-dashed), $\beta=7$ (dotted), $\beta=5$ (dot-dashed), 
$\beta=3$ (dashed), $\beta=1$ (solid).} 
\label{f:pia2}
\end{figure}

 
As in the previous section, we have evaluated the 
maximum interaction time $\tau ^{\star}$ that provides an 
output state---in the signal and idler modes---whose overlap 
with the state predicted by the parametric approximation is larger than 
$99\%$. The time $\tau ^{\star}$ for vacuum input and number input 
$|n\equiv  1,n\equiv 1\rangle$ is plotted as a function of the pump intensity 
in Fig. \ref{f:pia1}a. The corresponding achievable two-mode squeezing 
parameter---the maximum argument $|\chi_M |$ in the operator 
(\ref{twb})---is represented in Fig. \ref{f:pia1}b. 
In Fig. \ref{f:pia2} we show the average photon number 
$\langle n\rangle_{\hbox{\scriptsize out}}$ of the output signal mode 
as a function of the interaction time, and for different values of the pump 
amplitude. Notice that the quantity $a^{\dag}a-b^{\dag}b$ is conserved, 
so that, for the considered input states, the idler mode has the same 
average photon number as the signal one. 
\par
\begin{figure}
\begin{center}
\epsfxsize=.49\textwidth\leavevmode\epsffile{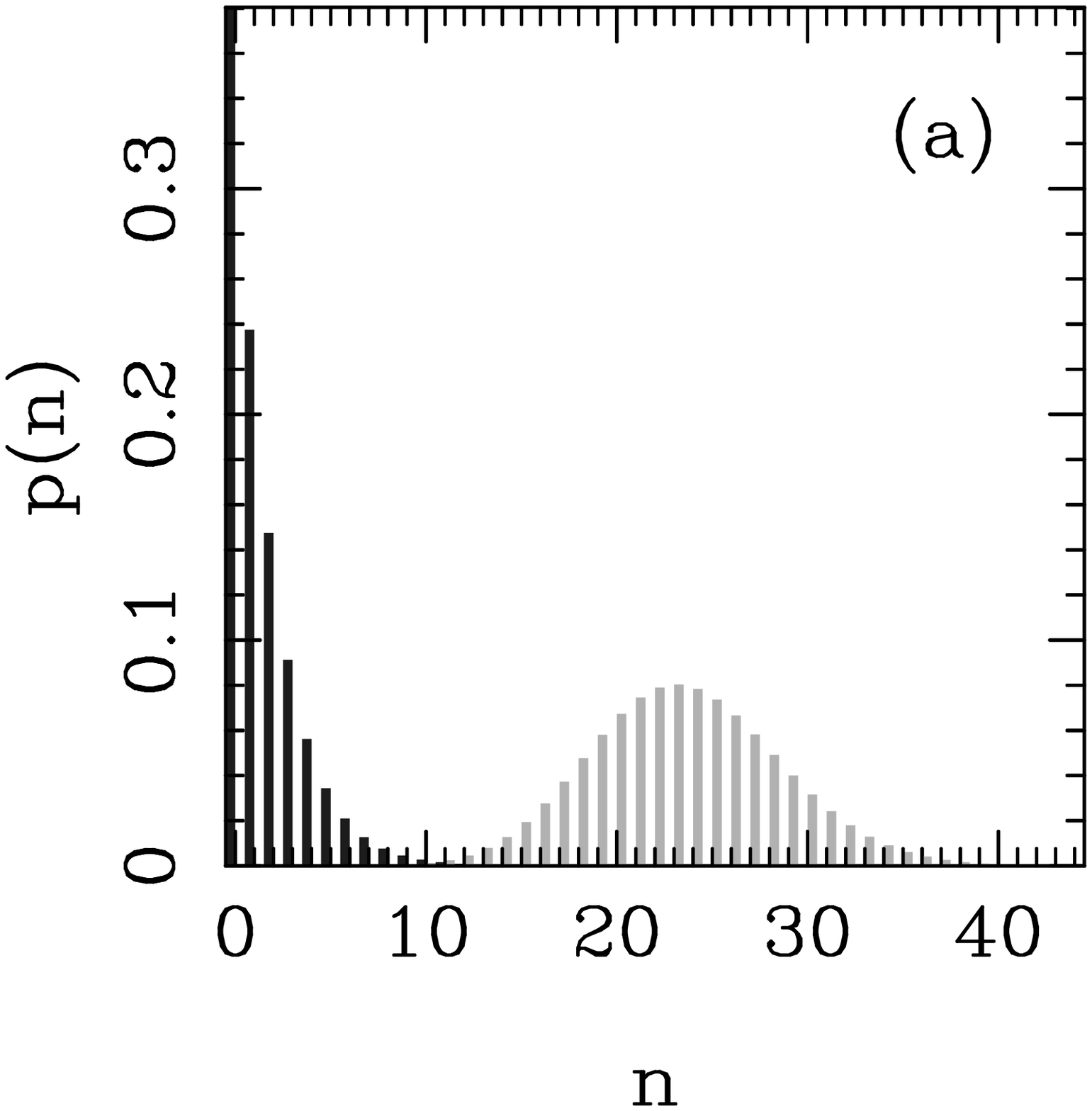}
\epsfxsize=.49\textwidth\leavevmode\epsffile{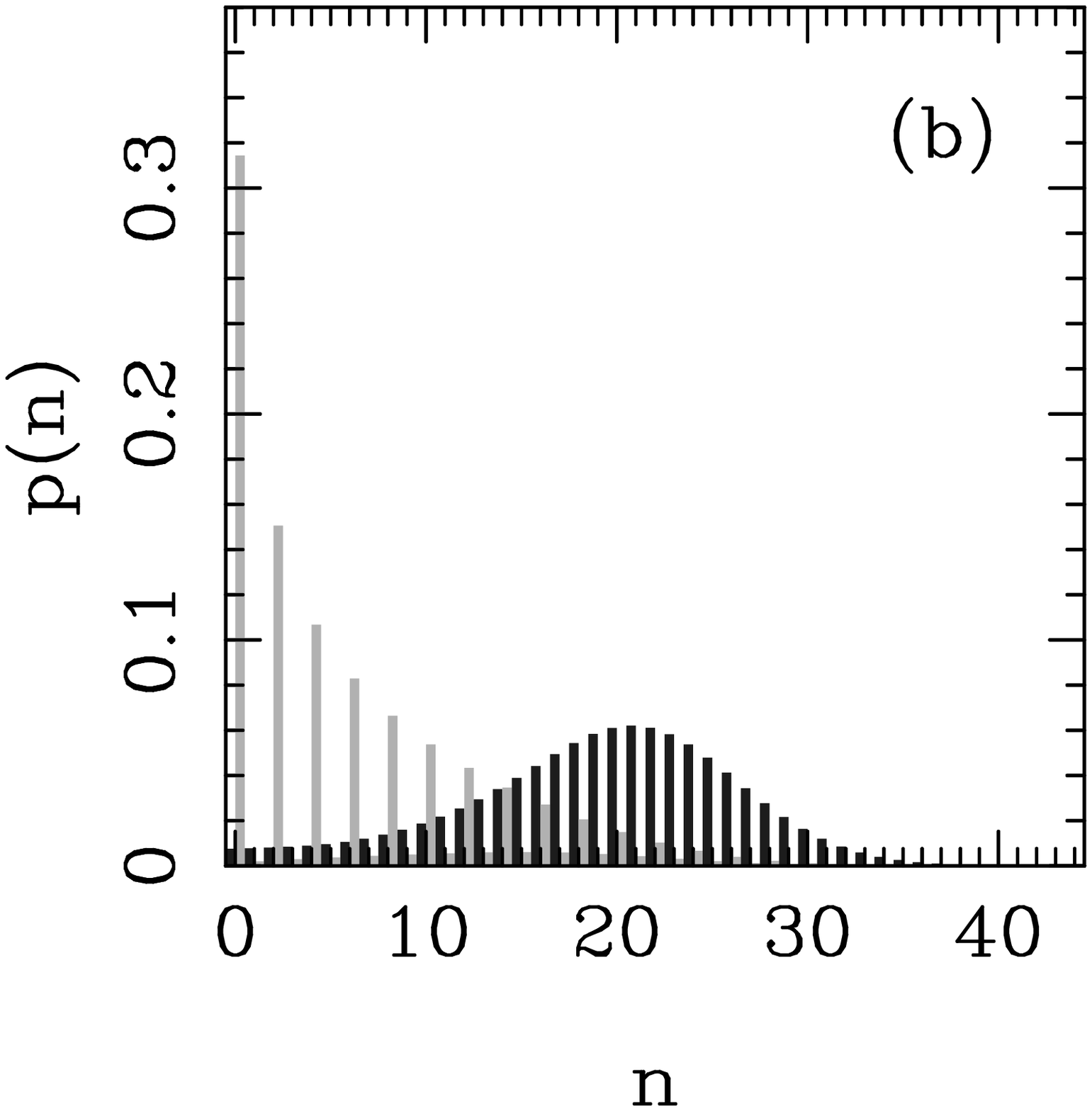}
\end{center}
\vskip .8cm
\caption{Photon number probabilities for both the signal and the pump
modes at the output of a nondegenerate parametric amplifier. The plots
refer to a situation in which the signal mode is initially in the
(two-mode) vacuum and the pump mode is excited to a coherent state
with amplitude $\beta=-5i$.  The interaction time is equal to
$\tau\equiv\tau^\star=0.214$ in (a), and to $\tau= 3 \tau^\star$ in
(b).  In the first case the parametric approximation well describes
the real interaction, which produces a twin-beam state with two-mode
squeezing parameter $\chi=0.789 $ corresponding to about $3.3$ output
photons.  On the other hand, parametric approximation does not hold in
the second case, as it can be easily recognized from the pump
squeezing.  The pump Fano factor in (b) is about $F=7.4$.}
\label{f:pia3}
\end{figure}

As shown for the degenerate case, here also the requirement of a strong pump 
is not peremptory, whereas the undepletion of the pump does not guarantee 
the validity of the parametric approximation. 
Again, it is the Fano factor $F$ of the pump after the 
interaction that well discriminates the working regimes of the amplifier. 
As long as $F\le 1.10$, the overlap between the states at the output 
and those predicted by the parametric approximation is larger than 
$99\%$. 
For interaction time longer than $\tau ^{\star}$, the pump mode 
reveals its quantum character, by showing oscillations in the 
number probability. 
This is illustrated in Fig. \ref{f:pia3}, where we report the photon number 
probabilities for both the signal and the pump modes in the case of vacuum 
input and pump amplitude equal to $|\beta |=5$, for different values of the 
interaction time. 
\section{Conclusions}
The quantum description of many optical devices is based on interaction 
Hamiltonians that couple different modes of radiation through the 
susceptibility tensor of the medium that supports the interaction.
The theoretical predictions about such interactions 
are usually drawn in the so-called parametric approximation, i.e. by treating 
the pump mode classically as a fixed $c$-number. 
Owing to such approximation, an analytical treatment is possible with the 
help of the factorization formulas for Lie algebras.
\par In this paper we have investigated the conditions under which the 
parametric approximation holds in the treatment of $\chi ^{(2)}$ 
nonlinear amplifiers, by resorting to the exact 
diagonalization of the their full Hamiltonians. We have explicitly 
compared the states evaluated by the exact evolution with those predicted 
by the parametric approximation, in terms of the overlap between such states. 
On one hand, we have shown that the regime of validity of the parametric 
approximation is very large, including also the case of weak pump 
with $1\div 10$ mean photon number. On the other, we have found that 
neither the condition of pump undepletion guarantees the goodness 
of the approximation, nor the condition of pump depletion signs its failure. 
We found that the degree of coherence of the pump {\em after} the 
interaction is a univocal parameter that discriminates the working regimes 
of the amplifiers. In terms of the pump Fano factor $F$ we found that a 
deviation smaller than $10\%$ guarantees an overlap larger than $99\%$ 
between the states predicted within the parametric approximation and those 
evaluated by the exact Hamiltonian. 
\par
For long interaction times the approximation breaks down, and 
the quantum character of the pump mode is revealed. 
Oscillations  in the pump number probability appear and, correspondingly, 
the Wigner function of the signal mode assumes negative values and resembles 
a Schr\"odinger-cat state.

\section*{Acknowledgment}
M. G. A. Paris would like to acknowledge the ``Francesco Somaini'' 
foundation for partial support.

\newpage
\newpage
\newpage
\newpage
\newpage
\newpage

\end{document}